\documentclass[12pt]{article}
\usepackage{epsfig}
\usepackage{cite}
\usepackage{axodraw}

\newcommand{\beq}{\begin{equation}}
\newcommand{\eeq}{\end{equation}}
\newcommand{\bea}{\begin{eqnarray}}
\newcommand{\eea}{\end{eqnarray}}
\newcommand{\nn}{\nonumber}

\def\eqn#1{Eq.~(\ref{#1})}

\def\sec#1{Sect.~{\ref{#1}}}
\def\app#1{Appendix~\ref{#1}}

\def\spa#1.#2{\left\langle#1\,#2\right\rangle}
\def\spb#1.#2{\left[#1\,#2\right]}
\def    \sapp#1#2#3#4{{\langle #1 | (#2+#3) |#4  \rangle} }

\def\bentarrow{\:\raisebox{1.3ex}{\rlap{$\vert$}}\!\longrightarrow}
\def\tr{\mathop{\rm tr}\nolimits}
\def\tree{{\rm tree}}
\def\oneloop{{1 \mbox{-} \rm loop}}
\def\flip{\hbox{flip}}
\def\eps{\epsilon}
\def\ord{{\cal O} }
\def\cA{{\cal A}}
\def\cP{{\cal P}}
\def\lra{\leftrightarrow}
\def\ellb{{\bar \ell}}

\def\ib{{\bar i}}
\def\qb{{\bar q}}
\def\d{{\rm d}}
\def\MS{$\overline{\rm MS}$}

\newcommand\sss{\scriptscriptstyle}
\newcommand\as{\alpha_{\sss S}} 
\newcommand\astwo{\alpha_{\sss S}^2} 
\newcommand\aem{\alpha_{\rm em}}
\newcommand\gs{g_{\sss S}}
\def\epem{e^+e^-}
  \newcommand{\ccaption}[2]{
    \begin{center}
    \parbox{0.85\textwidth}{
      \caption[#1]{\small{\it{#2}}}
      }
    \end{center}
    }

\makeatletter
\def\@eqnnum{\hbox{\reset@font\rm(\theequation)}}
\let\make@eqnnum=\@eqnnum %
\def\eqnum#1{\dec@eqnnum \global\def\make@eqnnum{\reset@font\rm(#1)}%
\def\@currentlabel{#1}%
}
\def\inc@eqnnum{\addtocounter{equation}{1}}
\def\dec@eqnnum{\addtocounter{equation}{-1}}
\@definecounter{equation}%
\@addtoreset{equation}{section} %
\def\theequation@prefix{{\thesection}.} %
\def\theequation{\theequation@prefix\arabic{equation}}%
\makeatother

\begin{document}

\begin{titlepage}

\hspace*{\fill}\parbox[t]{3.5cm}{
YITP-SB-00-75\\
GEF-TH-7/2000\\
hep-ph/0011368\\
}

\vspace{2cm}
\begin{center}
{\Large\bf QCD radiative corrections to $\gamma^*\gamma^* \to$ 
hadrons}\\
\vspace{2.cm}

{\large
{M. Cacciari$^1$, V. Del Duca$^2$, S. Frixione$^3$ and
Z. Tr\'ocs\'anyi$^4$\footnote{Sz\'echenyi fellow of the Hungarian
Ministry of Education}}
}

\vspace{.2cm}
{$^1$\sl C.N. Yang Institute for Theoretical Physics\\
State University of New York\\
Stony Brook, NY 11794-3840}

\vspace{.2cm}
{$^2$\sl I.N.F.N., Sezione di Torino\\
via P. Giuria, 1\ \ 10125 Torino, Italy}\\

\vspace{.2cm}
{$^3$\sl I.N.F.N., Sezione di Genova\\
via Dodecaneso, 33\ \ 16146 Genova, Italy}\\

\vspace{.2cm}
{$^4$\sl University of Debrecen and\\
Institute of Nuclear Research of the Hungarian Academy of Sciences\\ 
H-4001 Debrecen, PO Box 51, Hungary}\\

\vspace{.5cm}

\begin{abstract}
We compute the order-$\as$ corrections to the total 
cross section and to jet rates for the process
$\epem\to\epem +$~hadrons, where the hadrons are produced 
through crossed-channel quark exchange in the hard scattering 
of two off-shell photons originating from the incoming leptons.
We use a next-to-leading order general-purpose partonic Monte Carlo 
event generator that allows the computation of a rate differential 
in the produced leptons and hadrons. We compare our results with the 
available experimental data for $\epem\to\epem +$~hadrons at LEP2.
\end{abstract}

\end{center}
 \vfil

\end{titlepage}

\section{Introduction}

In measurements, and related theoretical analyses, of scaling violations
of the $F_2$ structure function~\cite{Aid:1995rk,Forte:1996xv},
of forward-jet production in DIS~\cite{Aid:1995we,Breitweg:1999ed,
Adloff:1999fa,Bartels:1996gr,Orr:1998tf}, and of dijet production at 
large rapidity intervals~\cite{Abachi:1996et,Abbott:2000ai,Mueller:1987ey,
DelDuca:1994mn,Stirling:1994zs,DelDuca:1995ng,Orr:1998hc}, several attempts 
have been made to detect a footprint of BFKL~\cite{Kuraev:1976ge,
Kuraev:1977fs,Balitsky:1978ic} evolution in hadronic cross sections.
Except for forward-jet production in DIS, where a full next-to-leading-order 
(NLO) calculation~\cite{Mirkes:1997jd} has proven itself insufficient to
describe the data, perhaps hinting toward corrections of BFKL type, 
none of the processes above shows any appreciable difference from a
standard perturbative-QCD behaviour, which allows us to describe them
in terms of a fixed-order expansion in $\as$ of the kernel cross
sections, complemented with the Altarelli-Parisi evolution of the parton
densities. However, in the processes above the hadronic nature of one or 
both of the incoming particles renders it difficult to disentangle an 
eventual BFKL signal from standard non-perturbative long-distance effects.

In order to overcome this problem, it was proposed in
Refs.~\cite{Mueller:1994rr,Mueller:1994jq} to consider the high-energy
scattering of two heavy quarkonia, since the transverse sizes of the quarkonia
are small enough to allow for the perturbative computation of their wave
function. At present, scattering of two heavy quarkonia is not feasible
experimentally. An increasingly popular alternative is the study of the
process
\beq
\gamma^* +\gamma^*\;\longrightarrow\; {\rm hadrons},
\label{processgg}
\eeq
at fixed photon virtualities $q_i^2=-Q_i^2 < 0$, and for large 
center-of-mass energies squared $W^2=(q_1+q_2)^2$,
with $q_i$ the momenta of the photons. The virtual photons play 
the same role as the quarkonia; they are colourless, and their
virtualities control their transverse sizes, which are roughly proportional 
to $1/\sqrt{Q_i^2}$, thus allowing for a completely perturbative
treatment. The virtuality of the photon is therefore
physically equivalent to the (squared) mass of the quarkonium; however,
while the mass of the quarkonium is fixed by nature, the virtuality
of the photon can be controlled by the experimental setup.
\begin{figure}[thb]
\centerline{
   \epsfig{figure=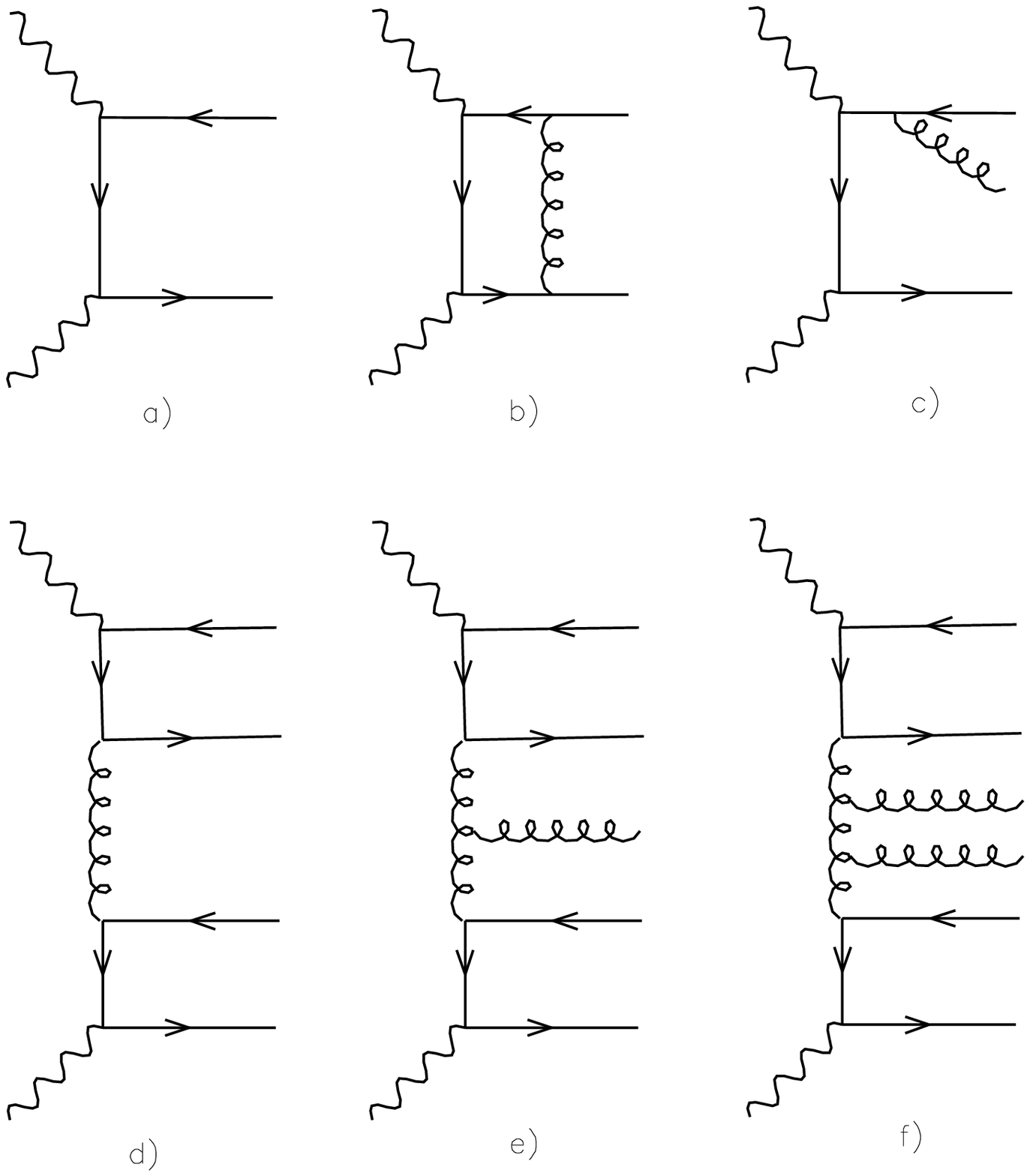,width=0.6\textwidth,clip=}
           }
\ccaption{}{\label{fig:diagr}
Sample of diagrams contributing to the production of hadrons
in the collision of two off-shell photons.
}
\end{figure}

In order to elucidate how the process in Eq.~(\ref{processgg}) may be
relevant to the BFKL dynamics, we expand the production rate associated to
Eq.~(\ref{processgg}) in $\as$ and illustrate in Fig.~\ref{fig:diagr}
some of the final-state configurations contributing to it.
Diagrams d), e) and f), plus all the diagrams obtained by emitting 
more and more gluons from the crossed-channel gluon,
are included in the BFKL dynamics: in fact, the BFKL theory assumes that any 
scattering process is dominated by gluon exchange in the crossed channel,
a result that holds exactly only asymptotically for large energies.
In the case at hand, if one considers the large-$W$ limit, diagrams with a
crossed-channel quark exchange, such as a), b) and c), are expected to give
a cross section behaving as
\beq
\sigma_{\gamma^*\gamma^*}\,\sim\,1/W^2,
\label{xsec}
\eeq
modulo logarithmic corrections,
while diagrams relevant to BFKL physics, such as d), e) and f), are
expected to give
\beq
\sigma_{\gamma^*\gamma^*}^{\sss BFKL}\,\sim\,
a_0+\sum_{j=1}^\infty a_j (\as L)^j + {\cal O}(\as(\as L)^j),
\label{BFKLxsec}
\eeq
where \mbox{$L=\log(W^2/\mu_{\sss\rm W}^2)$} is a ``large'' logarithm, and 
all subleading logarithmic terms are indicated with 
${\cal O}(\as(\as L)^j)$; the quantity $\mu_{\sss\rm W}^2$ is a
mass scale squared, typically of the order of the crossed-channel momentum
transfer and/or of the photon virtualities.
By comparing Eqs.~(\ref{xsec}) and (\ref{BFKLxsec}), it is clear that
the latter will dominate over the former in the asymptotic energy region
$W\to\infty$. Thus, testing the BFKL predictions in an ideal world would
be quite straightforward: the data relevant to the process in 
Eq.~(\ref{processgg}) for large $W$ values would be compared to 
the theoretical predictions for $\sigma_{\gamma^*\gamma^*}^{\sss BFKL}$.

However, things are not so simple when comparing the theory to the data
of a realistic experimental set-up. Firstly, at current collider energies
$\sigma_{\gamma^*\gamma^*}$ is not safely negligible, and must be
taken into proper account. For this
reason, one usually {\it subtracts} the theoretical predictions for
$\sigma_{\gamma^*\gamma^*}$ from the data, and then compares the results
obtained in this way to the predictions for
$\sigma_{\gamma^*\gamma^*}^{\sss BFKL}$, which have been obtained in the
high-energy limit in Refs.~\cite{Bartels:1996ke,Brodsky:1997sd,
Balitsky:1978ic}. Unfortunately, at present
only the leading order (LO) contribution to $\sigma_{\gamma^*\gamma^*}$ 
(diagram a) in Fig.~\ref{fig:diagr}) has been
considered~\cite{Vermaseren:1983cz,Gunion:1986mc}. 
Diagrams such as b) and c) have been neglected so far; these diagrams
represent the first non-trivial QCD corrections to the process in
Eq.~(\ref{processgg}): as we know from other processes in hadron physics, they
might give rise to a sizable enhancement with respect to the LO cross
section. We shall denote these contributions as NLO corrections, although
effectively of leading order in $\as$. The aim of this work is to compute the
NLO corrections, in order to assess whether a full NLO calculation of the
total cross section suffices to describe the data. As a by-product, we shall
also be able to give our predictions for one- and two-jet cross sections, with
the jets having a non-trivial internal structure. In the case of dijet
observables, we shall choose the two jets with the largest rapidity interval
in the event, in accordance with the Mueller-Navelet
analysis~\cite{Mueller:1987ey}. The total and jet cross sections are computed
by using a general-purpose parton generator, developed specifically for this
work.\footnote{The code can be obtained upon request.}

A second remark concerns the data relevant to the
process in Eq.~(\ref{processgg}). The easiest way to access this 
process is through the reaction,
\beq
\begin{array}{rcl}
e^+ + e^- & \longrightarrow & e^+ + e^- + \underbrace{\gamma^* + \gamma^*} \\
 &  & \phantom{e^+ + e^- + \gamma^*\:}\bentarrow {\rm hadrons} ;
\end{array}
\label{processee}
\eeq
namely, one considers $\epem$ collisions, selecting those events in
which the incoming leptons produce two photons which
eventually initiate the hard 
scattering that produces the hadrons. However, it is clear that the 
process in Eq.~(\ref{processee}) is non physical; rather, it has to be 
understood as a shorthand notation for a subset of Feynman diagrams 
contributing to the process that is actually observed,
\beq
e^+ +e^-\,\longrightarrow\,
e^+ +e^- + {\rm hadrons}.
\label{fullproc}
\eeq
Other contributions to the process in Eq.~(\ref{fullproc}) are, for example, 
those in which the incoming $\epem$ pair annihilates into a photon or a $Z$
boson, eventually producing the hadrons and a lepton pair, or those in which 
one (or both) of the two photons in Eq.~(\ref{processee}) is replaced
by a $Z$ boson. However, it is not difficult to devise a set of cuts such
that the process in Eq.~(\ref{processee}) gives the only non-negligible
contribution to the process in Eq.~(\ref{fullproc}). One can tag 
both of the outgoing leptons, and retain only those events (thus
termed {\em double-tag events}) in which the scattering angles 
of the leptons are small: in such a way, the contamination due 
to annihilation processes is safely negligible.
Furthermore, small-angle tagging also guarantees that the photon
virtualities are never too large (at LEP2, one typically measures
$Q_i^2={\cal O}(10$~GeV$^2$)); therefore, the contributions from processes
in which a photon is replaced by a $Z$ boson are also negligible.
Thus, it is not difficult to extract the cross section of the 
process \mbox{$\gamma^*\gamma^* \to$~hadrons} from the data relevant 
to the process in Eq.~(\ref{fullproc}). Double-tag events have in
fact been studied by the CERN L3 and OPAL Collaborations, at various $\epem$ 
center-of-mass energies ($\sqrt{S} =$ 91 and 183 GeV~\cite{Acciarri:1999ix}, 
189 GeV~\cite{opal} and 189-202 GeV~\cite{L3,L3ph2k,opal2000}).

With this in mind, we computed the cross section for the
process in Eq.~(\ref{processee}), rather than that relevant to the
process in Eq.~(\ref{processgg}); as it should be clear from the
previous discussion, the two are strictly equivalent from a physical
point of view. However, the former can be more easily related to 
the experimental analyses; in fact, our code outputs the momenta
of both the final-state partons and the leptons. The reader should
keep in mind that the study of QED radiative corrections shall not be
considered in this paper; in particular, we shall not address the
problem of a proper treatment of the initial state radiation, which is 
rather important on the experimental side, and is not fully understood 
yet for what concerns double-tag events. 

The outline of the paper is the following: in \sec{sec:rate} we explain
how the LO and NLO production rates are computed, giving some details
on the simplifications possible in the present case as compared to
other NLO calculations. Then, in \sec{sec:results} we present
phenomenologically relevant results: LO and NLO rates are computed 
for total, inclusive jet and dijet cross sections at LEP2 and at a 
possible configuration for a Next Linear Collider (NLC). In doing this,
we discuss the possible choices of mass scales entering the electromagnetic 
and strong running couplings, and analyse how the NLO rates depend on 
variations of the strong scale. Finally, we compare our results with the 
available data for the total cross section at LEP2. In \sec{sec:conclude} 
we draw our conclusions. A few useful formulae are reported in the appendices.

\section{Production rates}
\label{sec:rate}

The computation of the NLO corrections to a hard scattering process is
by now a rather standard procedure, since algorithms exist that are
universal (that is, process independent), and applicable to any number
of final state partons. The role of these algorithms is to combine
in a physically sensible way the virtual and the real contributions, 
that are unphysical and divergent upon loop and phase-space integrations.
The information on the hard process basically enter 
only in the computation of the matrix elements. In our case,
one needs to compute the amplitude of the process
\mbox{$\epem\to\epem q\bar{q}$} at one loop, and of the process
\mbox{$\epem\to\epem q\bar{q}g$} at the tree level. Fortunately, these
results are easily obtained from existing literature (notice that we
assume the quarks to be massless; in Sect.~\ref{sec:results} we shall 
comment further on this choice).
As a preliminary step, we need also to consider the process
\mbox{$\epem\to\epem q\bar{q}$} at the tree level, which gives the
LO contribution, first computed in Ref.~\cite{Vermaseren:1983cz,Gunion:1986mc},
and that we analyse in the following subsection.

\subsection{The LO matrix elements}
\label{sec:lorate}

In order to evaluate the matrix element relevant to the process
\beq
\begin{array}{rcl}
e^+ + e^- & \longrightarrow & e^+ + e^- + \underbrace{\gamma^* + \gamma^*} \\
 &  & \phantom{e^+ + e^- + \gamma^*\:}\bentarrow q+\qb\:,
\end{array}
\label{reaction}
\eeq
we use the helicity amplitudes for the
$\qb q \to \gamma^*\gamma^*  \to \ell \ellb \, \ellb' \ell'$ process,
with all the particles taken as outgoing. The scattering amplitude is 
(see Fig.~\ref{diagrams})
\beq
\cA_6(1_q,2_\qb;3_\ell,4_\ellb,5_{\ellb'},6_{\ell'}) = 4 e^4 Q_f^2
A_6(1,2;3,4,5,6)\, ,
\label{A6}
\eeq
with $e Q_f$ the electromagnetic charge of the quark $q$ of flavour $f$,
and where labels $\{1,2\}$ refer to the quark pair, while $\{3,4\}$ and 
$\{5,6\}$ denote the lepton pairs. The sub-amplitude $A_6$ depends only 
on the momenta and helicities of the external particles (we also point
out that Eq.~(\ref{A6}) is valid to any order in $\as$). At tree 
level, the sub-amplitude $A_6$ for any helicity configuration is
given in terms of a single function $a_6$,
\beq
A^{\tree}_6(1,2;3,4,5,6) = a_6(1,2;3,4,5,6) + a_6(1,2;6,5,4,3)\,.
\label{a6treesym}
\eeq
In terms of spinor products, currents and kinematic invariants as defined
in \app{sec:appb}, the explicit form of the function $a_6$ for the
$(1^-,2^+;3^-,4^+,5^+,6^-)$ configuration is the following
\cite{Dixon:1998py}:
\beq
a_6(1^-,2^+;3^-,4^+,5^+,6^-) = 
i \, { \spa1.3\spb2.5 \sapp6254 \over s_{34}\,s_{56}\,t_{134} } 
\: .\label{treea6}
\eeq
The function $a_6$ is odd under the flip symmetry
\beq
\flip : \qquad
1\leftrightarrow 2\,, \hskip 0.4 cm  
3\leftrightarrow 5\,, \hskip 0.4 cm  
4\leftrightarrow 6\,, \hskip 0.4 cm 
\spa{a}.{b} \leftrightarrow \spb{a}.{b} \, .\label{FlipSymmetry}
\eeq
In addition, there is a reflection symmetry on the quark line such that
\beq
a_6(1^+,2^-;3^-,4^+,5^+,6^-) = a_6(2^-,1^+;6^-,5^+,4^+,3^-)\, .\label{refl}
\eeq
Thus, the other quark-helicity configuration can be obtained by reflection,
which amounts to exchanging the labels 1 and 2 in \eqn{a6treesym}.
The other lepton-helicity configurations are obtained by exchanging the 
labels 3 and 4 and/or 5 and 6 in \eqn{treea6}.
\begin{figure}[t]
\begin{center}
\begin{minipage}{5cm}
\begin{picture}(100,100)(0,0)
\Line                ( 54.03, 87.26)( 20.03, 87.26)         
\ArrowLine           ( 54.03, 87.26)( 82.03,107.26)         
\ArrowLine           ( 54.03, 25.26)( 20.03, 25.26)         
\Line                ( 54.03, 25.26)( 82.03,  5.26)         
\Photon              ( 54.03, 87.26)( 66.03, 70.26){3}{4}   
\Photon              ( 54.03, 25.26)( 66.03, 42.26){3}{4}   
\Line                ( 93.03, 70.26)( 66.03, 70.26)         
\ArrowLine           ( 66.03, 42.26)( 66.03, 70.26)         
\Line                ( 66.03, 42.26)( 93.03, 42.26)         
\Text                ( 23.03, 93.26)[l]{ $4$ }
\Text                ( 20.03, 80.26)[l]{ $\to$ }
\Text                ( 20.03, 32.26)[l]{ $\to$ }
\Text                (  0.03, 87.26)[l]{ $e^-$ }
\Text                (  0.03, 25.26)[l]{ $e^+$ }
\Text                ( 84.03,110.26)[l]{ $e^-$ }
\Text                ( 84.03,  2.26)[l]{ $e^+$ }
\Text                ( 97.03, 70.26)[l]{ $q$ }
\Text                ( 97.03, 43.26)[l]{ $\bar q$ }
\Text                ( 60.03,104.26)[l]{ $3$ }
\Text                ( 24.03, 19.26)[l]{ $6$ }
\Text                ( 60.03,  8.26)[l]{ $5$ }
\Text                ( 76.03, 77.26)[l]{ $1$ }
\Text                ( 76.03, 35.26)[l]{ $2$ }
\Text                ( 40.00,-15.00)[l]{a)}
\end{picture}
\end{minipage}
\begin{minipage}{5cm}
\begin{picture}(100,100)(0,0)
\Line                ( 54.03, 87.26)( 20.03, 87.26)         
\ArrowLine           ( 54.03, 87.26)( 82.03,107.26)         
\ArrowLine           ( 54.03, 25.26)( 20.03, 25.26)         
\Line                ( 54.03, 25.26)( 82.03,  5.26)         
\Photon              ( 54.03, 87.26)( 66.03, 70.26){3}{4}   
\Photon              ( 54.03, 25.26)( 66.03, 42.26){3}{4}   
\Line                ( 93.03, 70.26)( 66.03, 70.26)         
\ArrowLine           ( 66.03, 42.26)( 66.03, 70.26)         
\Line                ( 66.03, 42.26)( 93.03, 42.26)         
\Text                ( 23.03, 93.26)[l]{ $4$ }
\Text                (  0.03, 87.26)[l]{ $e^-$ }
\Text                (  0.03, 25.26)[l]{ $e^+$ }
\Text                ( 84.03,110.26)[l]{ $e^-$ }
\Text                ( 84.03,  2.26)[l]{ $e^+$ }
\Text                ( 97.03, 70.26)[l]{ $q$ }
\Text                ( 97.03, 43.26)[l]{ $\bar q$ }
\Text                ( 60.03,104.26)[l]{ $3$ }
\Text                ( 24.03, 19.26)[l]{ $6$ }
\Text                ( 60.03,  8.26)[l]{ $5$ }
\Text                ( 76.03, 77.26)[l]{ $1$ }
\Text                ( 76.03, 35.26)[l]{ $2$ }
\Text                ( 40.00,-15.00)[l]{b)}
\Gluon               ( 78.26, 70.26)(78.26,42.26){3}{4}
\end{picture}
\end{minipage}
\begin{minipage}{5cm}
\begin{picture}(100,100)(0,0)
\Line                ( 54.03, 87.26)( 20.03, 87.26)         
\ArrowLine           ( 54.03, 87.26)( 82.03,107.26)         
\ArrowLine           ( 54.03, 25.26)( 20.03, 25.26)         
\Line                ( 54.03, 25.26)( 82.03,  5.26)         
\Photon              ( 54.03, 87.26)( 66.03, 70.26){3}{4}   
\Photon              ( 54.03, 25.26)( 66.03, 42.26){3}{4}   
\Line                ( 93.03, 70.26)( 66.03, 70.26)         
\ArrowLine           ( 66.03, 42.26)( 66.03, 70.26)         
\Line                ( 66.03, 42.26)( 93.03, 42.26)         
\Text                ( 23.03, 93.26)[l]{ $4$ }
\Text                (  0.03, 87.26)[l]{ $e^-$ }
\Text                (  0.03, 25.26)[l]{ $e^+$ }
\Text                ( 84.03,110.26)[l]{ $e^-$ }
\Text                ( 84.03,  2.26)[l]{ $e^+$ }
\Text                ( 97.03, 70.26)[l]{ $q$ }
\Text                ( 97.03, 43.26)[l]{ $\bar q$ }
\Text                ( 60.03,104.26)[l]{ $3$ }
\Text                ( 24.03, 19.26)[l]{ $6$ }
\Text                ( 60.03,  8.26)[l]{ $5$ }
\Text                ( 76.03, 77.26)[l]{ $1$ }
\Text                ( 76.03, 35.26)[l]{ $2$ }
\Text                ( 40.00,-15.00)[l]{c)}
\Gluon               ( 72.03, 70.26)(87.26,56.26){3}{2}
\Text                ( 88.03, 56.26)[l]{ $g$ }
\Text                ( 68, 56.5)[l]{ $7$ }
\end{picture}
\end{minipage}
\end{center}
\vspace{.5cm}
\ccaption{}{\label{diagrams} 
Sample of diagrams contributing to Eq.~(\ref{processee}), obtained
by dressing diagrams a), b) and c) of Fig.~\ref{fig:diagr} with
external lepton legs. The particle labelling scheme, as used
in Sect.~\ref{sec:rate}, is also shown.
}
\end{figure}

In crossing to the physical region, we choose 4 as the incoming
electron and 6 as the incoming positron. For a fixed lepton-helicity
configuration, e.g.\ $(3_\ell^-,4_\ellb^+,5_{\ellb'}^+,6_{\ell'}^-)$,
the production rate is obtained by summing over the quark-helicity
configurations,\footnote{In eqs.~(\ref{LOcross}), (\ref{NLOrealcross}),
and~(\ref{NLOvirtcross}) it is understood that the amplitudes are crossed 
into the physical channel.}
\bea
&&
\label{LOcross}
\d\sigma(3_\ell^-,4_\ellb^+,5_{\ellb'}^+,6_{\ell'}^-) =
{1\over 2S} \d\cP_{2+2}(p_1,p_2,p_3,p_5;p_4+p_6)
\,(4\pi\aem )^4 \Bigl( \sum_f Q_f^4 \Bigr)\,16 N_c
\\ \nn && \qquad\qquad\qquad\qquad\,
\times \left[ |A^{\tree}_6(1^-,2^+;3^-,4^+,5^+,6^-)|^2 + 
|A^{\tree}_6(2^-,1^+;3^-,4^+,5^+,6^-)|^2 \right]\,,
\eea
where $S=(p_4+p_6)^2$, $\d\cP_{2+2}$ is the phase space for the 
final-state quark pair and lepton pair (see \eqn{phasespace}), and
\beq
\sum_f Q_f^4 = Q_u^4 n_u + Q_d^4 n_d\,,
\eeq
with $Q_u=2/3$, $Q_d=-1/3$ and $n_{u(d)}$ being the number of
up(down)-type quarks.

\subsection{The NLO matrix elements}
\label{sec:nlorate}

At NLO, we must consider the real corrections due to the emission of a 
gluon off the quark line, $\gamma^*\gamma^* \to q\bar q g$, and the 
one-loop corrections to $\gamma^*\gamma^* \to q\bar q$. For the gluon
emission, we can use the tree amplitude
$ \qb q g \to \gamma^*\gamma^*  \to \ell \ellb \, \ellb' \ell' $,
\beq
\cA^\tree_7(1_q,2_\qb;3_\ell,4_\ellb,5_{\ellb'},6_{\ell'};7_g) = 4 e^4 Q_f^2
\gs \lambda_{i_1}^{~\ib_2}
A^\tree_7(1,2;3,4,5,6;7)\,.\label{A7tree}
\eeq
The colour subamplitude $A^\tree_7$ can again be written in terms of a
single function $a_7$
\beq
A^{\tree}_7(1,2;3,4,5,6;7) = a_7(1,2;3,4,5,6;7) + a_7(1,2;6,5,4,3;7)\, .
\label{a7treesym}
\eeq
For the configuration $(1^-,2^+;3^-,4^+,5^+,6^-;7^+)$ it
reads~\cite{Dixon:1998py}
\bea
\lefteqn{a_7(1,2;3,4,5,6;7) = } 
\\ \nn &&
i \, {\spa1.3 \over \spa1.7 s_{34}\,s_{56}\,t_{134}} \left[ 
   {\spa1.3\spb3.4\spb2.5\sapp6257 \over t_{256}}
 + {\sapp6134 \sapp1275 \over \spa7.2} \right]\,.\label{treea7}
\eea
The same configuration with a negative-helicity gluon is obtained
by applying the $-$flip operation of Eq.~(\ref{FlipSymmetry}).

For the lepton-helicity configuration
$(3_\ell^-,4_\ellb^+,5_{\ellb'}^+,6_{\ell'}^-)$, the production rate is 
obtained by summing over the quark and gluon helicity configurations,
\bea
&&
\label{NLOrealcross}
\!\!\!\!\!\!\!\!\!\!\!\!\!\!\!\!
\d\sigma_r(3_\ell^-,4_\ellb^+,5_{\ellb'}^+,6_{\ell'}^-) = 
{1\over 2S} \d\cP_{3+2}(p_1,p_2,p_7,p_3,p_5;p_4+p_6)\,
(4\pi\aem)^4\Bigl( \sum_f Q_f^4 \Bigr)\,16 (N_c^2-1)
\\ \nn && \qquad\qquad\qquad
\times 4\pi\as  
\left[ \left( |A^{\tree}_7(1^-,2^+;3^-,4^+,5^+,6^-;7^+)|^2 + \flip 
\right) + (1 \lra 2) \right] \, ,
\eea
where $\d\cP_{3+2}$ is the phase space for the three QCD partons ---
the $q\bar{q}$ pair and the gluon --- and the lepton pair.

In order to compute the one-loop corrections to 
$\gamma^*\gamma^* \to q\bar q$, we can use the
one-loop amplitude for $\qb q \to \gamma^*\gamma^*  \to \ell \ellb \, 
\ellb' \ell' $, which we can extract from the one-loop amplitude for
$e^+e^- \to q_1\bar q_1 q_2\bar q_2$~\cite{Bern:1998sc} by replacing
the quark-gluon vertex factor $\gs\lambda^a$ with the quark-photon vertex
factor $\sqrt{2}\, e Q_f$\footnote{The factor $\sqrt{2}$ is due to the 
$\tr(\lambda^a \lambda^b) = \delta^{ab}$ normalization, where
$\lambda^a$ are the generators of the SU(3) group in the fundamental
representation.}.
The unrenormalized one-loop amplitude is given by \eqn{A6} with $A_6
\to A^\oneloop_6$ substitution, where
\beq
A^\oneloop_6 = {N_c^2-1 \over N_c} \gs^2 c_{\Gamma} \left(
A^{\tree}_6 V + i F \right)\, .
\eeq
$A^{\tree}_6$ is given in \eqn{a6treesym} and the prefactor $c_{\Gamma}$ is
\beq
c_{\Gamma} = {1\over 
(4\pi)^{2-\epsilon}}\, {\Gamma(1+\epsilon)\,
\Gamma^2(1-\epsilon)\over \Gamma(1-2\epsilon)}\, .\label{cgam}
\eeq
The universal divergent piece $V$, in the dimensional 
reduction~\cite{Siegel:1979wq,Capper:1980ns} scheme
or four-dimensional helicity scheme~\cite{Bern:1992aq} used to compute the
one-loop amplitude, reads
\beq
V =   - {1\over \eps^2} \left( {\mu^2 \over -s_{12}}\right) ^\eps 
  - {3\over 2 \eps} \left( {\mu^2 \over -s_{12}}\right) ^\eps   - 4\, .
\label{Vab}
\eeq
The one-loop charge renormalization UV counterterm to $A^\oneloop_6$ is
zero, due to the electric-charge conservation.  The finite piece $F$ is
obtained from Eq.~(12.11) of Ref.~\cite{Bern:1998sc} by performing on
it the relabeling $\{1,2,3,4,5,6\} \to \{5,6,2,1,3,4\}$.  For the
lepton-helicity configuration
$(3_\ell^-,4_\ellb^+,5_{\ellb'}^+,6_{\ell'}^-)$, the one-loop
production rate is then,
\bea
\label{NLOvirtcross}
\lefteqn{\d\sigma_v(3_\ell^-,4_\ellb^+,5_{\ellb'}^+,6_{\ell'}^-) = 
{1\over 2S}\d\cP_{2+2}(p_1,p_2,p_3,p_5;p_4+p_6)\,
(4\pi\aem)^4 \Bigl( \sum_f Q_f^4 \Bigr)\,16 N_c}
\\ \nn &&
\times
\left\{ 2 {\rm Re} \left[ A^{\tree}_6(1^-,2^+;3^-,4^+,5^+,6^-)^* 
A^{\oneloop}_6(1^-,2^+;3^-,4^+,5^+,6^-) \right] + (1 \lra 2) \right\}\,.
\eea
In Eqs.~(\ref{NLOrealcross}) and (\ref{NLOvirtcross}),
the other three lepton-helicity configurations are obtained by
permuting the lepton labels as described in \sec{sec:lorate}.
The unpolarised rate is given by averaging the fixed-helicity rates 
over the four lepton-helicity configurations.
In order to obtain the correct cross section in conventional dimensional
regularization, we have to add the term~\cite{Catani:1997pk}
\beq
\label{DRtoCDR}
-\frac{\as C_F}{2\pi}\,\d\sigma(3_\ell^-,4_\ellb^+,5_{\ellb'}^+,6_{\ell'}^-)
\eeq
to the right hand side of \eqn{NLOvirtcross}, where
$\d\sigma(3_\ell^-,4_\ellb^+,5_{\ellb'}^+,6_{\ell'}^-)$ is given in
\eqn{LOcross}.

\subsection{From matrix elements to physical observables}
\label{sec:phasespace}

\begin{figure}[thb]
\centerline{
   \epsfig{figure=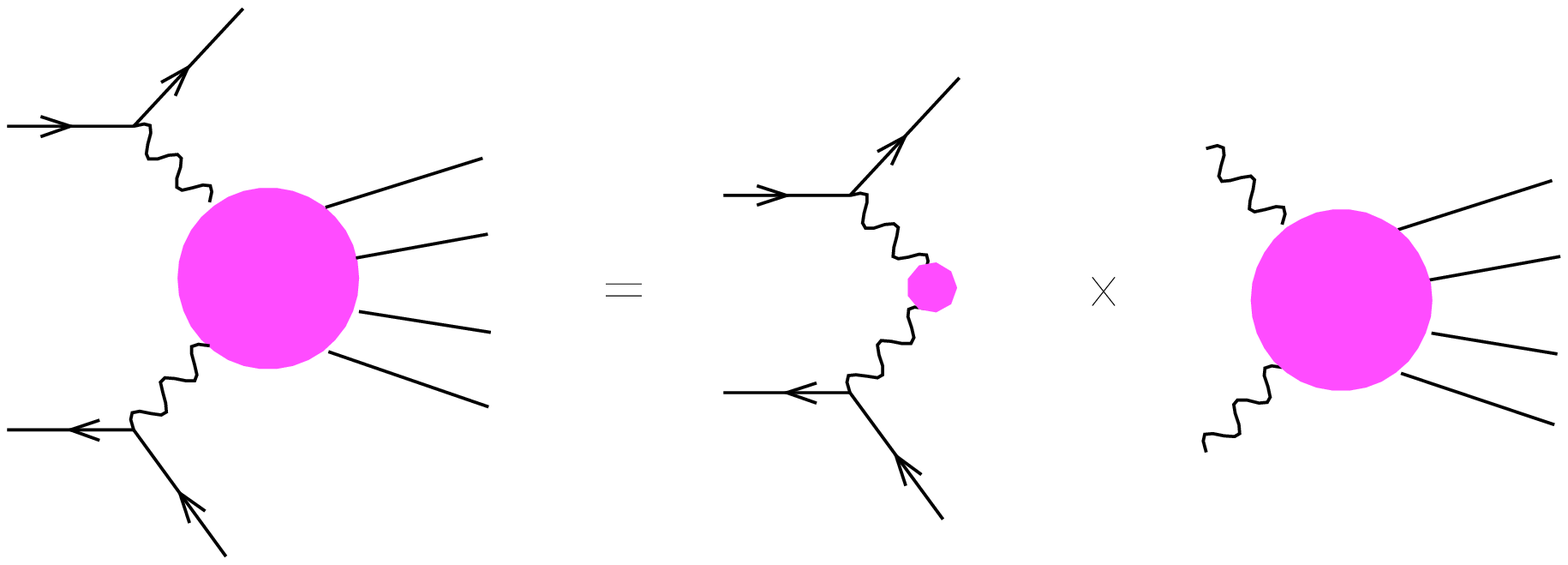,width=0.7\textwidth,clip=}
           }
\ccaption{}{\label{fig:decomp}
Decomposition of the phase space for the process
$\epem\to\epem +$~hadrons.
}
\end{figure}

Having the matrix elements at disposal, one can plug them into one's
preferred NLO algorithm, and obtain physical results. Our case can
however be greatly simplified in a preliminary stage; in fact, the
incoming and outgoing leptons do not participate in the hard
scattering, that is initiated by the two virtual photons. Formally,
the simplification goes through a suitable decomposition of the phase
space. This is achieved by writing the phase space of two leptons 
plus $n$ partons (in our case, $n=2$ or $n=3$ for the one-loop or the 
tree-level amplitudes respectively) as follows,
\beq
\d\cP_{n+2}(k_1,\dots,k_n,p_{\ell_1},p_{\ell_2};
p_{\ell_1}^\prime+p_{\ell_2}^\prime) =
\d\Gamma(p_{\ell_1},p_{\ell_2})\,
\d\cP_n(k_1,\dots,k_n;q_1+q_2)\,,
\label{phspdec}
\eeq
where $p_{\ell_i}^\prime$ ($p_{\ell_i}$) are the momenta of the incoming
(outgoing) leptons, $q_i=p_{\ell_i}^\prime-p_{\ell_i}$, $k_i$ are the 
momenta of the outgoing partons, and we used the standard definition
of the phase space of $m$ particles,
\beq
\d\cP_{m}(r_1,\dots,r_m;R)=
(2\pi)^4 \, \delta^4 \left(R-\sum_{i=1}^m r_i \right)\,
\prod_{i=1}^m \frac{\d^3 r_i}{(2\pi)^3 2r_i^0}\,.
\label{phasespace}
\eeq
The decomposition in Eq.~(\ref{phspdec}) is represented pictorially
in Fig.~\ref{fig:decomp}: the lepton sector
communicates with the hadron sector only through the photon momenta
$q_i$. 
It has to be stressed that we changed the labelling convention for the
momenta with respect to the previous subsections; while the former
labelling rendered it easy to write the matrix elements in terms of helicity
amplitudes, the present one, which we shall adopt from now on, is more
transparent from the physical point of view. We can get back to the 
labelling of the previous subsections with the following identifications,
\beq
p_{\ell_1}^\prime\to p_4,\;\;
p_{\ell_2}^\prime\to p_6,\;\;
p_{\ell_1}\to p_3,\;\;
p_{\ell_2}\to p_5,\;\;
k_1\to p_1,\;\;
k_2\to p_2,\;\;
k_3\to p_7.
\eeq
Eqs.~(\ref{phspdec}) and~(\ref{phasespace}) implicitly define $\d\Gamma$,
and we get
\beq
\d\Gamma(p_{\ell_1},p_{\ell_2}) =
{\d^3 p_{\ell_1}\over (2\pi)^3 2 p_{\ell_1}^0}\,
{\d^3 p_{\ell_2}\over (2\pi)^3 2 p_{\ell_2}^0}\,.
\label{dPhi}
\eeq
From Eqs.~(\ref{phasespace}) and~(\ref{dPhi}), we see that
both terms in the right-hand side of Eq.~(\ref{phspdec}) have a
Lorentz-invariant expression. We exploit it to re-write 
Eq.~(\ref{dPhi}) in the center-of-mass frame of the incoming $\epem$ 
pair,
\beq
\d\Gamma=\frac{1}{4(2\pi)^6 S}\,\d Q_1^2\,\d Q_2^2\,\d E_1\,\d E_2
\,\d\varphi\,\d\bar{\varphi},
\label{dPhiepem}
\eeq
where $\varphi$ and $\bar{\varphi}$ are two generic azimuthal angles, one
of which, say $\varphi$ to be definite, can be interpreted as the
angle between the two outgoing leptons; $E_i$ are the energies
of the outgoing leptons in the center-of-mass frame of the incoming $\epem$ 
pair. The strategy of the computation should now be clear: although we 
compute the cross section for the process in Eq.~(\ref{processee}), the hard
process we deal with at NLO is effectively that of Eq.~(\ref{processgg}).
Thanks to the decomposition in Eq.~(\ref{phspdec}), we have a $2\to n$
phase space which is formally identical to that one gets as a starting
point of any NLO algorithm. Thus, we can safely adopt one of the
existing NLO algorithms, and study the process of Eq.~(\ref{processgg})
in the $\gamma^*\gamma^*$ center-of-mass frame, without any reference
to the incoming or outgoing leptons. This amounts to a non-trivial
simplification, since the complexity of the numerical computations
at NLO is known to grow rapidly with the number of particles involved
in the hard scattering. Of course, the information on the lepton
momenta is entering somewhere, in particular in the matrix elements;
to take this fact into account, we proceed in two steps, still using 
Fig.~\ref{fig:decomp} as a guide. We start by generating the full kinematical
configuration of the outgoing leptons, using Eq.~(\ref{dPhiepem}).
In doing this, we also get the photon momenta, and therefore we know
how to boost from the $\epem$ to the $\gamma^*\gamma^*$ center-of-mass
frame. Then, we boost the lepton momenta to the $\gamma^*\gamma^*$ 
center-of-mass frame, where we generate the remaining (parton) momenta,
according to the phase space $\d\cP_n$; at this stage, we can perform
all the manipulations required by the NLO algorithm. More details
on the final-state kinematics, including bounds on the phase-space
variables, can be found in Appendix A.

Following the procedure outlined above, we constructed a code 
capable of predicting, to NLO accuracy, any infrared-safe quantity 
constructed with up to three partons (plus two leptons) in the final 
state. We stress that the code is {\em not} of a parton-shower type,
and should actually be regarded as a Monte Carlo integrator; however,
exactly like in the case of parton-shower Monte Carlo event generators, 
it allows us 
to easily implement realistic experimental cuts and to obtain binned
differential distributions for all sorts of variables and jet definitions.
The code is based upon the NLO algorithm of 
refs.~\cite{Frixione:1996ms,Frixione:1997np}, and it is a suitable
modification of one of the codes presented in Ref.~\cite{Frixione:1997np}.
A few technicalities concerning the code are given in Appendix A.

\section{Results}
\label{sec:results}

In this section, we present results of phenomenological relevance
obtained with the code mentioned above. Before assessing the effect 
of the NLO corrections and comparing our predictions to data, we 
discuss the choice of the scales entering the electromagnetic and the
strong running couplings which appear in the amplitudes. Although 
arguments exist on the choice of an ``optimal'' scale, there is no
rigorous theorem that prescribes such a choice. 
Thus, we shall choose the reference scale on the ground of some physical 
motivations; we stress, however, that alternative choices are possible,
and we shall explore a few of them.

\subsection{Scale choices}
\label{sec:scale}
As far as $\aem$ is concerned, we have made the choice of evolving
it on an event-by-event basis to the scales set by the virtualities of
the exchanged photons; hence, we replace the Thomson value $\alpha_0
\simeq 1/137$ by $\aem(Q_i^2)$. This choice better describes the 
effective strength at which the electromagnetic interaction takes place. 
Notice that we treat independently the two photon legs: thus, in the
formulae relevant to the cross sections, $\alpha_{\rm em}^4$ has to be 
understood as \mbox{$\alpha_{\rm em}^2(Q_1^2)\alpha_{\rm em}^2(Q_2^2)$}.

An analysis of the differential distributions reveals that, for
a typical experimental set-up used at LEP2, defined more precisely 
below, the mean $Q^2$ values fall in the 14--17 GeV$^2$ range. At this
scale the strength of the electromagnetic interaction is
increased by about 3\% with respect to the Thomson value. Since
$\aem$ enters at the fourth power in our squared matrix
elements, the 3\,\% increase in $\aem$
translates into an increase of more than 10\% in the cross
section. Some uncertainty is of course implicit in this number: both the
scheme for $\aem$ evolution (we used one-loop \MS~running) and the
precise scale value will affect the final result by a few per cent.

A similar problem is faced when considering the strong coupling $\as$, 
and it is solved in the same way: we define a default scale $\mu_0$ so 
as to match the order of magnitude of the (inverse of the) interaction 
range:
\beq
\mu_0^2 = \frac{Q_1^2+Q_2^2}{2} +
\left(\frac{k_{1{\sss T}} + k_{2{\sss T}} + k_{3{\sss T}}}{2}\right)^2 \: .
\label{defscale}
\eeq
The renormalization scale $\mu$ entering $\as$ will eventually be set 
equal to $\mu_0$ as a default value, and equal to $\mu_0/2$ or $2\mu_0$ when
studying the scale dependence of the cross section. In Eq.~(\ref{defscale}),
the $k_{i{\sss T}}$ are the transverse energies of the outgoing quarks and, 
for three-particle events, the emitted gluon. Since the hard process is 
initiated by the two virtual photons, the proper frame to study its
properties is the $\gamma^*\gamma^*$ center-of-mass one. Therefore,
when talking about transverse energies, whether in a total cross 
section or in a jet reconstruction algorithm, this frame will be always 
understood. This is in fact quite similar to what happens in DIS, where 
the Breit frame is used. Finally, we point out that the term in
parentheses in Eq.~(\ref{defscale}) is, event-by-event, half of the 
total transverse energy, which is a measurable quantity; at LO, it
coincides with the transverse energy of the jets, in the case in
which a jet cross section is considered.

We evolve $\as$ to next-to-leading log accuracy, 
with $\as(M_{\sss Z})=0.1181$~\cite{Groom:2000in} (in $\overline{{\rm MS}}$ 
at two loops and with five flavours, this implies 
$\Lambda_{\overline{{\rm MS}}}^{(5)} = 0.2275$ GeV).

We also considered a different choice with respect to that in
Eq.~(\ref{defscale}); namely, we used $\sqrt{Q_1^2 Q_2^2}$ instead
of $(Q_1^2+Q_2^2)/2$ in Eq.~(\ref{defscale}). Only {\em very} minor 
differences (much smaller than 1\%, i.e., not noticeable on the scale of the 
plots shown in what follows) were found. This is easily understood since 
the two scales coincide when $Q_1^2=Q_2^2$, and, as we verified
explicitly both at the LO and at the NLO, the dominant contribution
to the cross section is just due to the region where the virtualities
of the two photons are approximately equal. 

A third possible choice for the default scale is
\beq
\bar{\mu}_0^2 = \frac{Q_1^2+Q_2^2}{2}\,.
\label{muscale}
\eeq
Strictly speaking, $\bar{\mu}_0$ is arguably better than $\mu_0$ when
studying fully inclusive quantities, while $\mu_0$ is clearly
recommended when, for examples, jets are reconstructed. Still,
having a tool such as an event generator, we stick to $\mu_0$ 
as our default choice also for fully inclusive observables. 
However, we also studied the effect of setting $\mu=\bar{\mu}_0$, 
and found only minor differences (of order of 1\%) at the level of
total cross sections. We shall comment further on the use of 
$\bar{\mu}_0$ in Sect.~\ref{sec:datacomp}.

\subsection{Numerical results}
\label{sec:numres}

We compared our LO result, obtained with fixed $\aem = \alpha_0$, 
to the massless limit of the JAMVG program of Ref.~\cite{Vermaseren:1983cz},
and found perfect agreement.

To study the effect of the NLO corrections, we used the experimental cuts
employed by the L3 Collaboration; any other physically sensible sets 
of cuts would lead to the same qualitative conclusions. The scattered 
electron and positron are required to have energy $E_{1,2}$ larger than 30 
GeV and scattering angle $\theta_{1,2}$ between 30 and 66 mrad. Furthermore, 
the variable $Y$, defined by
\beq
Y=\log\frac{y_1 y_2 S}{\sqrt{Q_1^2 Q_2^2}} \equiv \log\frac{S}{S_0} ,
\label{YD}
\eeq
is required to lie between 2 and 7 ($y_i$ are defined in Appendix A,
where a discussion on the properties of $Y$ can also be found). The cross 
sections have been evaluated at $\sqrt{S} = 200$ GeV, including up to
five massless flavours.
\begin{table}[thp]
\begin{center}
\begin{tabular}{|c|c|c|}
\hline
LO, fixed $\aem=\alpha_0$ & LO, running $\aem(Q^2)$ & 
NLO, running $\aem(Q^2)$ \\
\hline
0.466   &  0.534  &  0.569$^{+0.006}_{-0.004}$ \\
\hline
\end{tabular}
\ccaption{}{\label{table1} 
Total cross section (in picobarns) within the 
L3 experimental cuts at LEP2 energy. The errors in the NLO column refer 
to the variation of the renormalization scale in the 
($\mu_0/2,2\mu_0$) range.
}
\end{center}
\end{table}

Within this set of cuts, the replacement of the
Thomson electromagnetic coupling $\alpha_0$
with the running one 
is found to increase our LO cross section by about 14\% 
(see Table~\ref{table1}),  in agreement
with the estimate given above. Such a non-negligible effect should of
course be included when comparing to experimental data.
Unless stated otherwise, all cross sections considered below
will be calculated with the running $\aem$.

Table~\ref{table1} also shows the effect of including the NLO
corrections calculated in this paper. They increase the LO
total cross section within the cuts applied by about 7 per cent.
This increase is of similar size as the $\as/\pi$ NLO correction to the
total hadronic cross section in electron-positron annihilation.
The numbers quoted as errors affecting the NLO result are
the differences between the cross sections obtained by choosing
$\mu=\mu_0/2, 2\mu_0$, and the cross section obtained by using
the default value, $\mu=\mu_0$. Thus, they should not be interpreted 
as statistical errors affecting our prediction, but rather as an
indication of the theoretical uncertainties due to the scale choice.
\begin{figure}[t]
\begin{center}
\begin{minipage}{\textwidth}
\epsfig{file=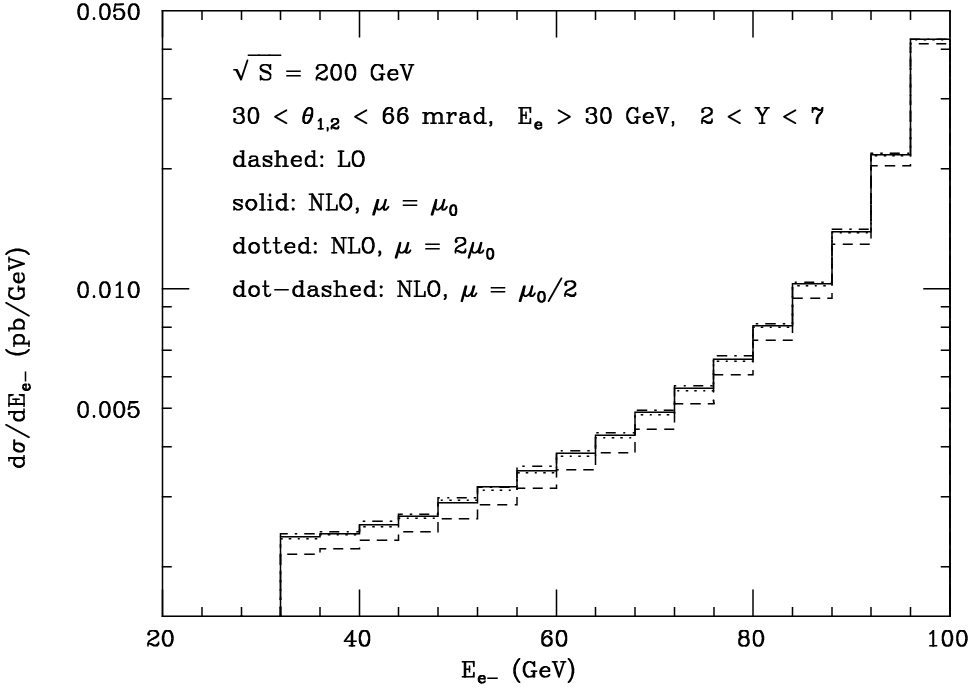,width=8cm}\hfill
\epsfig{file=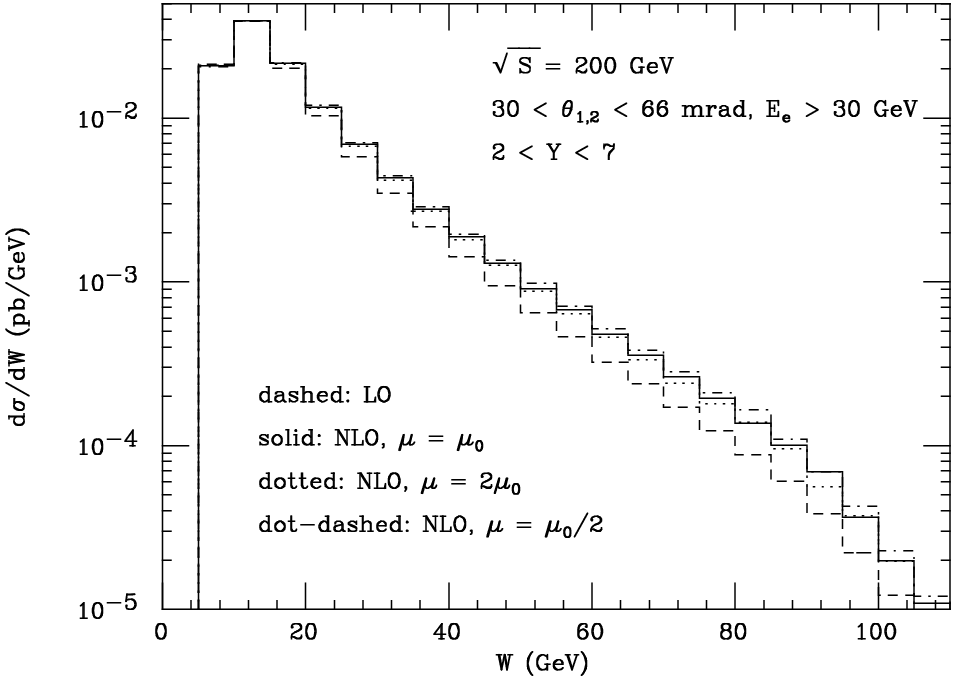,width=8cm}
\vspace{1cm}
\end{minipage}
\begin{minipage}{\textwidth}
\epsfig{file=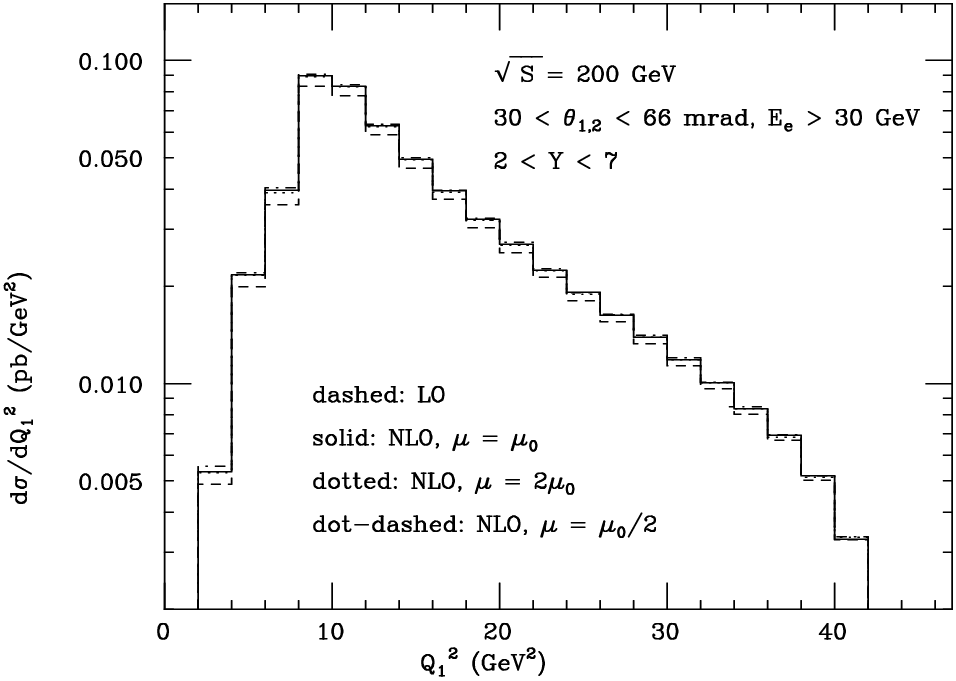,width=8cm}\hfill
\epsfig{file=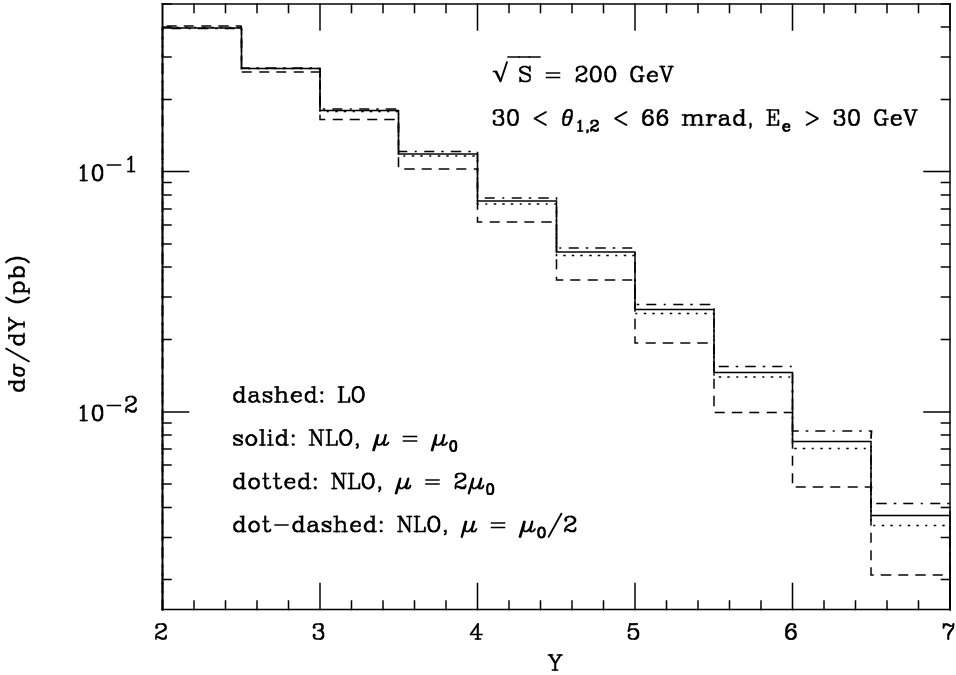,width=8cm}
\end{minipage}
\ccaption{}{\label{fig2} 
Differential cross sections within the L3
experimental cuts at LEP2 energy. Shown are the LO prediction (dashed line) 
and the NLO ones corresponding to three different choices for the
renormalization scale $\mu$.
}
\end{center}
\end{figure}

A better grasp on the effect of the radiative corrections can of course
be obtained by studying various differential distributions. 
Figure~\ref{fig2} shows such distributions for various observables of
experimental interest: the energy of the outgoing electron 
$E_{e^-}(\equiv E_1)$, the hadronic invariant mass $W$, the photon 
virtuality $Q_1^2$, and $Y$ as defined in Eq.~(\ref{YD}). 
In each plot the leading order curve and the three next-to-leading order
ones referring to the three choices ($\mu_0/2,\mu_0,2\mu_0$) 
of the renormalization scale $\mu$ are presented.

The uncertainty related to $\mu$ can be seen to be always smaller than 
the net effect of including the NLO corrections. It is actually quite
difficult to distinguish between the three NLO results (except in the
large-$W$ and $Y$ regions), the relative difference between them being
of about 1\% or less for the total cross section. This also implies that
the shape of the distribution is basically independent of $\mu$. 

As for the effect of the NLO corrections themselves, we see that, apart
from slightly increasing the cross section, they induce visible shape
modifications in at least two cases: both the $W$ and the $Y$
distributions become harder after the inclusion of radiative
corrections, their effect changing from almost nil at the left edge of
the plots to a more than 50\% increase at the right one.

Using similar experimental cuts we can also analyse the effect
of radiative corrections on jet distributions. We define the jets by
means of a $k_{\sss T}$ clustering algorithm~\cite{Catani:1993hr},
in the version formulated in Ref.~\cite{Ellis:1993tq}. We set the
jet-resolution parameter $D=1$ (see Ref.~\cite{Ellis:1993tq}). 
Contrary to the cuts previously used, here we do not impose an upper
limit on $Y$, only requiring $Y>2$. We consider single-inclusive jet
and dijet cross sections. In the latter case, we select the jets
by imposing a $E_{\sss T}>14$ GeV cut on the transverse energy of the most
energetic jet and requiring $E_{\sss T}>10$ GeV for at least another
jet. We adopt different transverse energy cuts on the two 
tagged jets in order to avoid the problems that arise in the
case in which such cuts are chosen to be equal, as discussed in
some details in Ref.~\cite{Frixione:1997ks}. Furthermore, as already
mentioned in the Introduction, in the case in which three jets are
present in the event, we take as tagged jets the two separated by the largest
rapidity interval. Finally, we shall only present results obtained
with $\mu=\mu_0$.
\begin{figure}[t]
\begin{center}
\epsfig{file=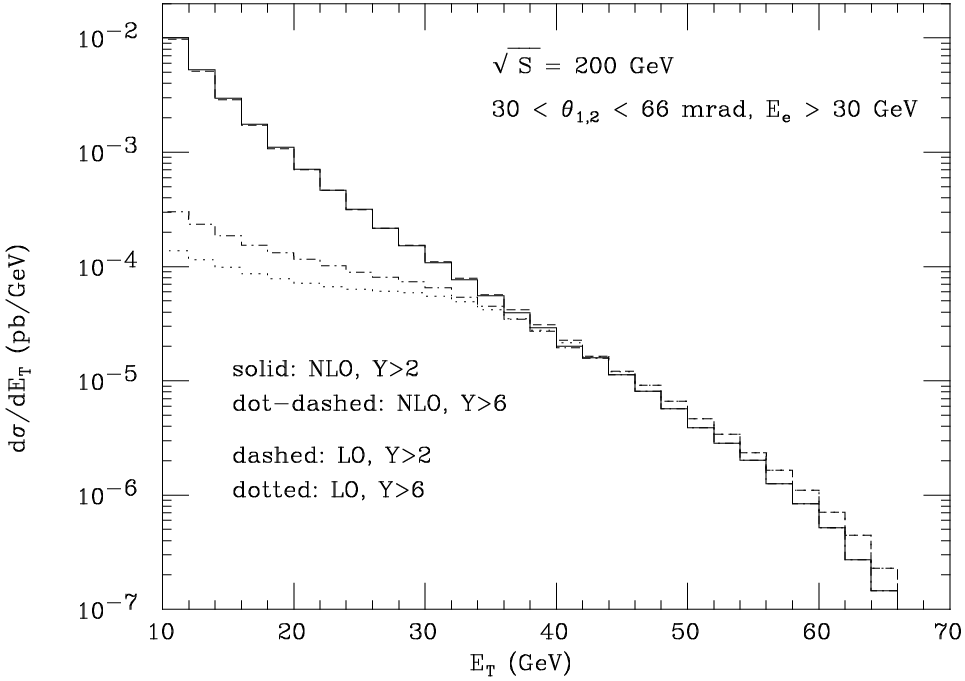,width=8cm}\hfill
\epsfig{file=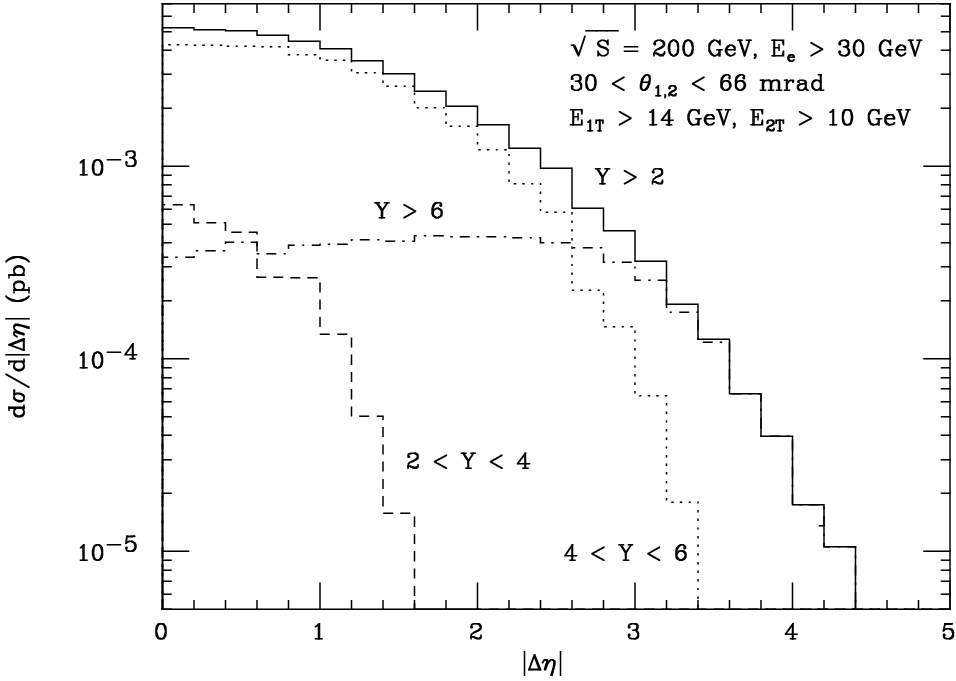,width=8cm}
\ccaption{}{\label{fig3} 
Differential distributions in jet events: transverse energy in 
single-inclusive jet production (left panel), and rapidity difference 
between the most forward/backward jets (right panel) in dijet production.
}
\end{center}
\end{figure}

In the left panel of Fig.~\ref{fig3} we show the transverse energy
distribution of single-inclusive jets, considering the cuts $Y>2$ 
and $Y>6$. The first striking feature of this observable is that the 
curves relevant to $Y>2$ and $Y>6$ coincide for $E_{\sss T}>40$~GeV.
This is so for the following reason: at the threshold (where the jets
are produced at zero rapidity), $W^2=4E_{\sss T}^2$; thus using
Eq.~(\ref{Yapproxdef}) with $E_{\sss T}=40$~GeV and $Q_1^2=Q_2^2=16$~GeV$^2$
(which is approximately the average virtuality within the current cuts), 
we get $Y=5.99$ (here, we identify $Y$ with $\overline{Y}$; see
Eq.~(\ref{YvsY})). Therefore, the region $2<Y<6$ simply does not
contribute to events with $E_{\sss T}>40$~GeV. On the other hand,
at $E_{\sss T}=40$~GeV, the two-photon system has just enough
energy, at $Y=6$, to produce the jets. Larger values of $Y$ do
not contribute much, since the $Y$ spectrum is very rapidly
falling at large $Y$'s (see Fig.~\ref{fig2}). When considering
larger transverse momenta, the situation is exactly the same.
We are therefore led to the conclusion that the tail of the 
$E_{\sss T}$ spectrum is dominated by threshold production,
and therefore cannot be reliably predicted by a fixed-order
computation, like ours; a resummation of large threshold 
logarithms is necessary. A signal that this is indeed the case is
reflected in the fact that the radiative corrections are negative 
in the tail. At smaller transverse energies the behaviour of
the radiative corrections displays a pattern similar to that
of total rates. For $Y>2$, NLO and LO results are very close
to each other; the larger $E_{\sss T}$, the more important the
contributions from threshold production. For $Y>6$, the radiative
corrections increase sizably the LO result; this is in agreement with
the behaviour of the $Y$ spectrum shown in Fig.~\ref{fig2}. The
increase is obviously related to the appearance of large logarithms
in the cross section, as it is always the case when two scales 
(here, the small $E_{\sss T}$ and the large $W$) are present.
We shall soon see that the large logarithms in the large-Y region are
indeed of BFKL type.

We also considered the transverse energies of the most forward
and most backward jet in dijet events, and found a pattern 
identical to that relevant to single-inclusive jet $E_{\sss T}$.
The reason is clear: even at small transverse energies, 
three-jet production is clearly disfavoured with respect
to dijet production; in the vast majority of the events,
there are just two hard jets recoiling against each other.

If we want to study jet production in a sensible way at fixed order,
we have therefore to consider observables which are as insensitive
as possible to threshold effects. From what we said above, such 
observables are possibly those that get the dominant contribution
from the small-$E_{\sss T}$ region. An example is given by rapidities.
In the right panel of Fig.~\ref{fig3}, we show the distributions in the
rapidity interval $\Delta\eta$ between the two tagged jets in dijet events, 
for various cuts on $Y$. In this case, only the NLO results are shown. We 
have verified that the radiative corrections give positive contributions
for all the regions in $Y$ considered, except for $2<Y<4$; in
this case, in fact, the energy of the two-photon system is so small 
that there is no way to get contributions away from the threshold.
The most interesting feature of this plot is that it shows that
the large-$Y$ and the large-$\Delta\eta$ regions select the same events,
as can be inferred from the fact that the distributions relevant to
$Y>2$ (solid line) and to $Y>6$ (dot-dashed line) exactly coincide
for $\Delta\eta>3.5$. This is actually the same behaviour we observe
in the case of the transverse energy distribution, but the underlying
physics is rather different. In fact, in this case we also get sizable
contributions away from the threshold; thus, at fixed $E_{\sss T}$,
part of the energy of the two-photon system contributes to the 
longitudinal degrees of freedom, and jets can be produced away from
the central region. Since we are in any case dominated by two-jet 
events, the rapidity difference between the two tagged jets can 
be easily estimated: $\Delta\eta\simeq \log(W^2/E_T^2)$.
Therefore, by using Eqs.~(\ref{Yapproxdef}) and~(\ref{YvsY}),
we get $Y\simeq \Delta\eta + \ln(E_T^2/\sqrt{Q_1^2Q_2^2})$.
We point out that the pattern displayed in Fig.~\ref{fig3} for large
$\Delta\eta$ does not depend upon the transverse momentum cuts: we
lowered these cuts down to 5~GeV, and found the same behaviour.
The large-$Y$ region is thus naturally suitable to study BFKL physics.
In addition, we note that the dijet cross section at NLO is rather small; 
therefore, with the integrated luminosities at LEP2 a sizeable number of 
dijet events would hint toward the importance of BFKL-type contributions.
Having clarified that the large $Y$ region is basically populated by events
characterised by two hard jets well separated in rapidity, we can follow
Ref.~\cite{Brodsky:1997sd}: we invert Eq.~(\ref{LvsY}) to get $c_{\sss Q}$, 
substituting $Y=6$ and identifying $L$ with the average $\Delta\eta$ 
corresponding to the cut $Y>6$ (in this way, we just make a choice for
the scale $\mu_{\sss\rm W}$ entering the BFKL logarithms; other choices
are of course possible, and all of them are equally good at the leading 
logarithm level). We have
\beq
\log c_{\sss Q}=Y-\langle\Delta\eta\rangle\simeq 4.32.
\label{logcq}
\eeq
By inspection of Eq.~(\ref{LvsY}), we see that, although $Y$ and the
BFKL logarithm $L$ coincide asymptotically, at LEP2 the difference
between the two is of the same order of $Y$, and thus cannot be neglected.
It seems therefore that LEP2 is quite far from probing the asymptotic BFKL 
region; it must be stressed, however, that the value given in Eq.~(\ref{logcq})
depends crucially on the assumptions made in Ref.~\cite{Brodsky:1997sd}.

It is presumed that a BFKL signature from double-tag hadronic
events would be observed at an hypothetical Next Linear Collider
much more easily than at LEP2. For this to be true, one actually needs
fairly small tagging angles, that allow to get relatively small values 
for the virtualities, with large $W$ values obtainable thanks to the large 
$\epem$ center-of-mass energy; in fact, it is argued~\cite{Bartels:1996ke}
that it would be desirable to tag the electrons down to 20-40 mrad.
We therefore studied the effect of NLO radiative corrections at
a NLC with $\sqrt{S} = 500$~GeV, requiring $E_{1,2} > 40$~GeV, $20 <
\theta_{1,2} < 70$~mrad and $Y > 2$; however, we point out that, at 
present, it seems unlikely that experiments at the NLC will reach 
such small values for the tagging angles.

The predicted total cross section within these cuts is found to be 
0.425~pb at leading order and 0.452~pb at next-to-leading order. The 
6\% increase is thus similar to the one found at LEP2. The same is 
true for the $Y$ distribution: for $Y<7$, that is in the range accessible 
both at LEP2 and at the NLC, the ratio of NLO over LO predictions
is to a very good extent the same in the two cases, getting as high
as 1.6 at $Y=6$. However, NLC within the cuts given above reaches
much larger value in $Y$ ($Y=11$), where the ratio of NLO over LO
gets to values of about 2.5. Finally, we verified that the pattern
shown in the right panel of Fig.~\ref{fig3} is reproduced also at
the NLC: the large-$Y$ and the large-$\Delta\eta$ regions are populated
by the same events. Clearly, as in the case of the $Y$ distribution,
the values of $\Delta\eta$ accessible at the NLC are larger than at
LEP2 (at NLC, $\Delta\eta<7.5$, for transverse energy cuts on jets
as given above).

The large NLO corrections that we find in the large-$Y$ region
at the NLC show that a calculation of the higher order effects
will be necessary in order to sensibly compare the theoretical 
predictions with the data, and eventually to extract evidence 
of BFKL dynamics from the latter.

\subsection{Comparisons with experimental data at LEP2}
\label{sec:datacomp}

\begin{table}[htb]
\begin{center}
\begin{tabular}{|c|c|c|c|}
\hline
\multicolumn{4}{|c|}{$d\sigma/dY$ (pb)~~~~~~~ $\sqrt{S} = 189 - 202$
GeV}\\
\hline
$\Delta Y$&L3 Data                     & LO    & NLO  \\
\hline
2.0 -- 2.5& 0.50 $\pm$ 0.07 $\pm$ 0.03 & 0.405 &0.396$^{+0.002}_{-0.002}$ \\
2.5 -- 3.5& 0.29 $\pm$ 0.03 $\pm$ 0.02 & 0.213 &0.225$^{+0.001}_{-0.002}$ \\
3.5 -- 5.0& 0.15 $\pm$ 0.02 $\pm$ 0.01 & 0.067&0.080$^{+0.002}_{-0.002}$\\
5.0 -- 7.0& 0.08 $\pm$ 0.01 $\pm$ 0.01 & 0.0091&0.0131$^{+0.0009}_{-0.0006}$\\
\hline
\hline
Total & 0.93 $\pm$ 0.05 $\pm$ 0.07  & 0.534  & 0.569$^{+0.006}_{-0.004}$ \\
\hline
\end{tabular}
\ccaption{}{\label{table2} 
The experimental cross section from L3 compared to leading and 
next-to-leading order predictions. The uncertainties in the NLO column
are related to variations of the renormalization scale.
}
\end{center}
\end{table}
The L3~\cite{Acciarri:1999ix,L3,L3ph2k} and OPAL~\cite{opal,opal2000} 
Collaborations have recently analysed data for hadron production in 
$\gamma^*\gamma^*$ collisions at an electron-positron center-of-mass 
energy around 200 GeV. In this Section, we aim at comparing these
data to our NLO results. We remind the reader that our predictions are 
all given at the {\em parton} level, as compared to the data that
are of course at the {\em hadron} level. 

\begin{figure}[t]
\begin{center}
\epsfig{file=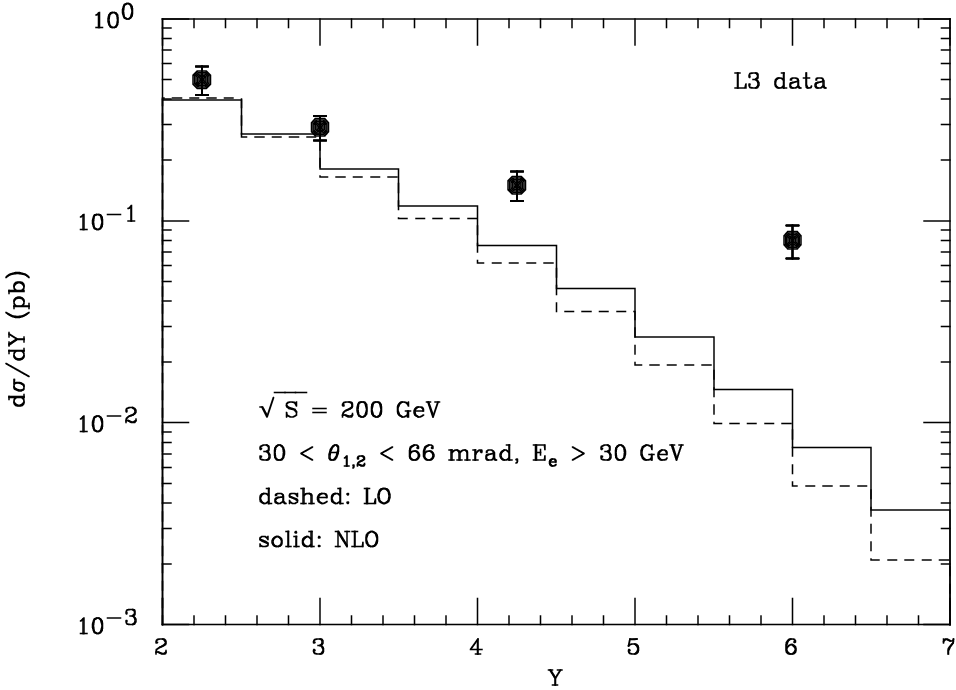,width=11cm}
\ccaption{}{\label{fig-l3}
Differential cross section with respect to $Y$ from the L3
Collaboration compared to leading and next-to-leading order
predictions. The data are taken at $\sqrt{S} = 189 - 202$~GeV. 
The theoretical simulation is always run at $\sqrt{S} = 200$ GeV.}
\end{center}
\end{figure}

L3 made use of the previously mentioned set of experimental cuts. The
cross section they find, as a function of $Y$, is reported in
Table~\ref{table2} and plotted in Fig.~\ref{fig-l3}. Table~\ref{table2} 
shows, in four different $Y$ bins, the experimental cross section compared
to our leading and next-to-leading order predictions, evaluated at $\sqrt{S}
= 200$~GeV. The same comparison is made in Fig.~\ref{fig-l3}: the data 
lie above the theory in the low-$Y$ region, and sizably overshoot the 
predictions in the large-$Y$ one. Thus we find a marked difference in 
shape between theory and data which, if confirmed, could be interpreted 
as the onset of important higher order effects, perhaps of BFKL type.
As can be seen from Table~\ref{table2}, the scale uncertainties
affecting our predictions are much smaller than the experimental
errors; in what follows, we shall therefore refrain from varying
the renormalization scale, setting it always equal to its default
value $\mu_0$. Also, the total cross section does tend to be higher 
than the predictions, as shown in Table~\ref{table2}. We remind the
reader (see Table~\ref{table1}) that the running of the electromagnetic 
coupling and the inclusion of the NLO corrections have raised the 
theoretical result from the 0.466 pb given by the massless leading 
order parton model with $\aem=\alpha_0$.

We also compared L3 data of Table~\ref{table2} to the predictions
obtained by choosing $\bar{\mu}_0$ as a reference scale (see 
Eq.~(\ref{muscale})). As remarked before, the effect on the total
cross section is rather small; however, our NLO predictions
for the two largest-$Y$ bins in Table~\ref{table2} get increased
by about 6\% and 15\% respectively. We are thus getting closer
to data, but still a very clear disagreement is seen between
theory and experiment.

We have also studied the effect of the finite mass of the outgoing
heavy quarks in the charm and bottom case, by comparing our results
with the ones obtained with the JAMVG~\cite{Vermaseren:1983cz} code.
Within the L3 set of cuts, such mass effects can be seen to decrease
the LO massless cross section by an amount of the order of  10-15\%.
One could in principle rescale the NLO result by this amount and get a
phenomenologically sensible prediction but, due to the lack of rigorousness
of this procedure, we shall always present our plots and numerical
results without such a correction.

The OPAL Collaboration has also recently presented
data~\cite{opal2000} taken at $\sqrt{S} = 189$ - 202 GeV, making use of
a slightly  different set of cuts: the tagged electron and positron
were required to have energies $E_{1,2} > 0.4 E_{beam}$ and angles $34 <
\theta_{1,2} < 55$ mrad. No cut on $Y$ is applied, but the
hadronic invariant mass $W$ is required to be larger than 5 GeV. 
Our simulation is run within these cuts at an $e^+e^-$ center-of-mass
energy corresponding to the luminosity-weighted average energy of the
OPAL data, i.e. $\sqrt{S}=194$~GeV.

\begin{table}[htb]
\begin{center}
\begin{tabular}{|c|c|c|c|}
\hline
\multicolumn{4}{|c|}{$d\sigma/d\overline{Y}$ (pb)~~~~~~~ $\sqrt{S} = 189$ - 202
GeV}\\
\hline
$\Delta \overline{Y}$&OPAL Data               & LO    & NLO  \\
\hline
0 -- 1& 0.055 $\pm$ 0.016 $^{+0.030}_{-0.020}$ & 0.068 & 0.062\\
1 -- 2& 0.118 $\pm$ 0.024 $^{+0.009}_{-0.024}$ & 0.140 & 0.133\\
2 -- 3& 0.123 $\pm$ 0.028 $^{+0.010}_{-0.011}$ & 0.090 & 0.093\\
3 -- 4& 0.070 $\pm$ 0.021 $^{+0.006}_{-0.015}$ & 0.043 & 0.049\\
4 -- 6& 0.028 $\pm$ 0.013 $^{+0.002}_{-0.012}$ & 0.011 & 0.014\\
\hline
\hline
Total & 0.40 $\pm$ 0.05 $\pm$ 0.05  & 0.364  & 0.365 \\
\hline
\end{tabular}
\ccaption{}{\label{table3} The experimental cross section from OPAL,
total and differential in $\overline{Y}$, compared to leading 
and next-to-leading order predictions.
}
\end{center}
\end{table}
Table~\ref{table3} compares the experimental results obtained with these 
cuts with our LO and NLO predictions. We can see the NLO corrections to be
extremely small. The prediction for the total cross section falls short of the
central OPAL result, but is well within the experimental error. Also shown 
in the same table is the differential distribution in the variable
$\overline{Y}$, defined in \eqn{Yapproxdef}, where a generally good 
agreement within errors can be observed. Given the large discrepancy
between theory and L3 data for this very same variable, it shall therefore
be of utmost importance to measure as accurately as possible the $Y$ spectrum,
in order to perform a precise study of effects beyond NLO (such as BFKL 
dynamics).

\begin{figure}[thp]
\begin{center}
\begin{minipage}{\textwidth}
\epsfig{file=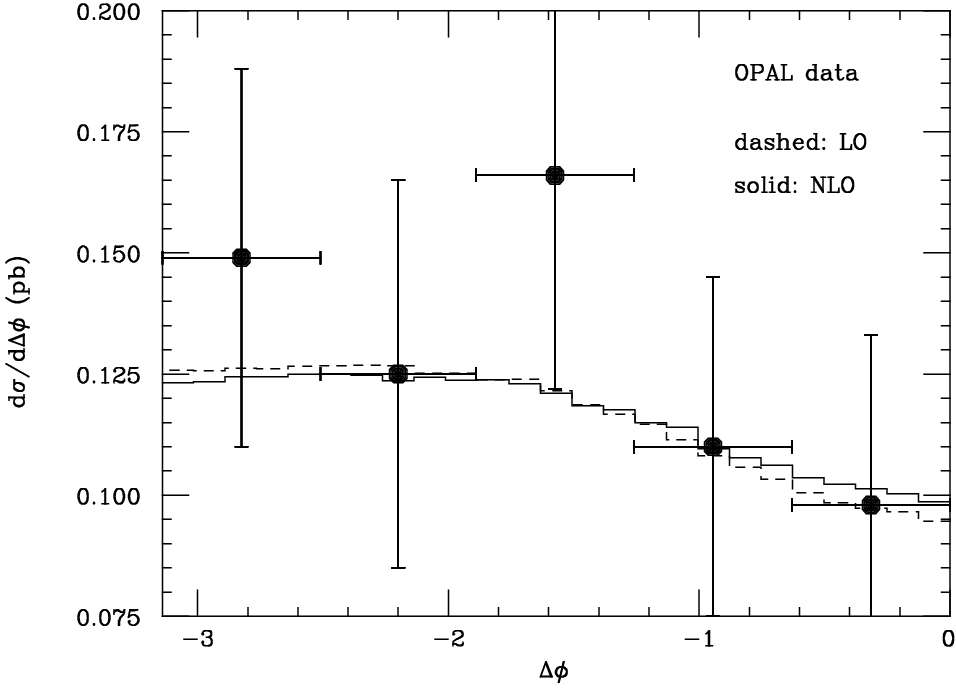,width=8cm}\hfill
\epsfig{file=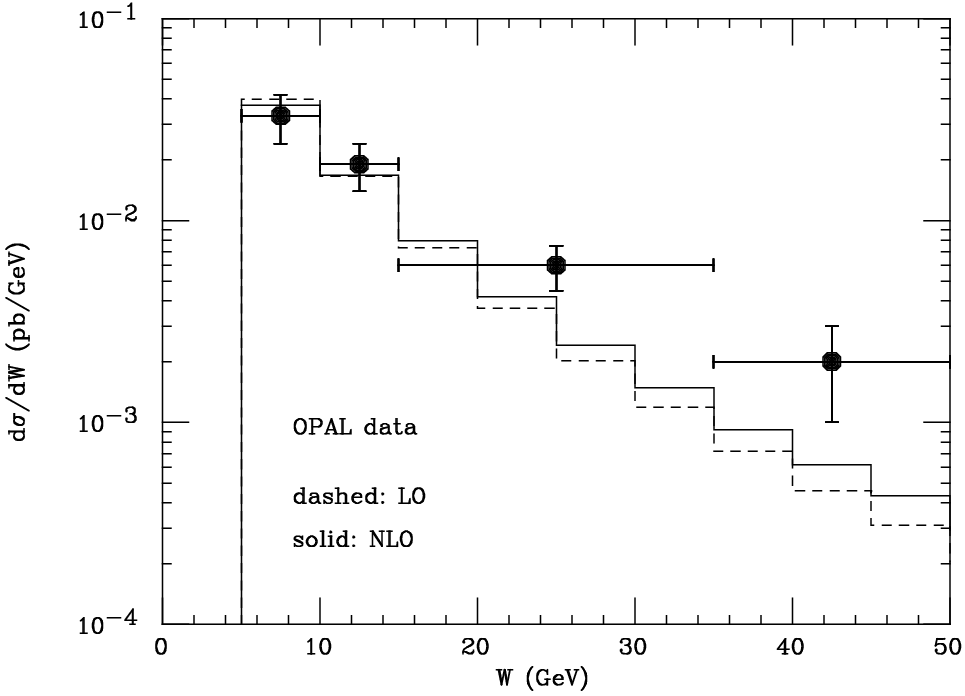,width=8cm}
\vspace{1cm}
\end{minipage}
\begin{minipage}{\textwidth}
\epsfig{file=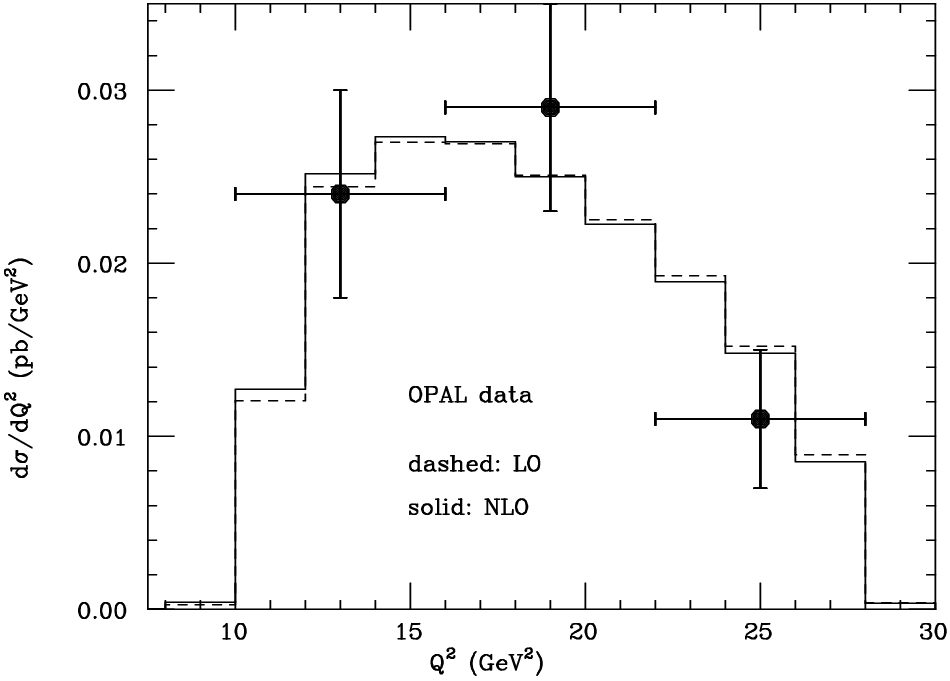,width=8cm}\hfill
\epsfig{file=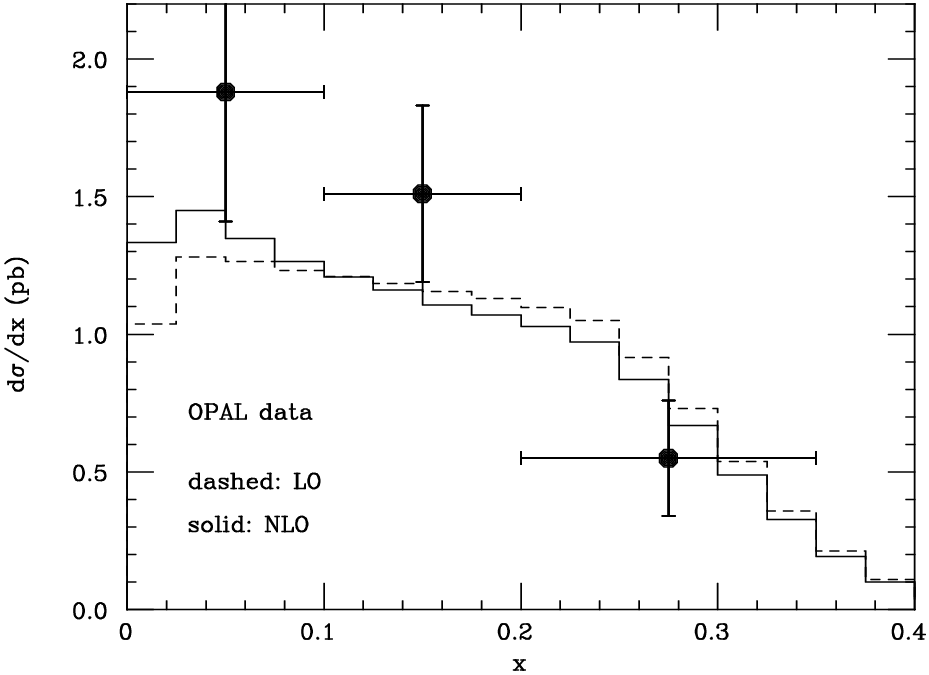,width=8cm}
\vspace{1cm}
\end{minipage}
\begin{minipage}{\textwidth}
\begin{center}
\epsfig{file=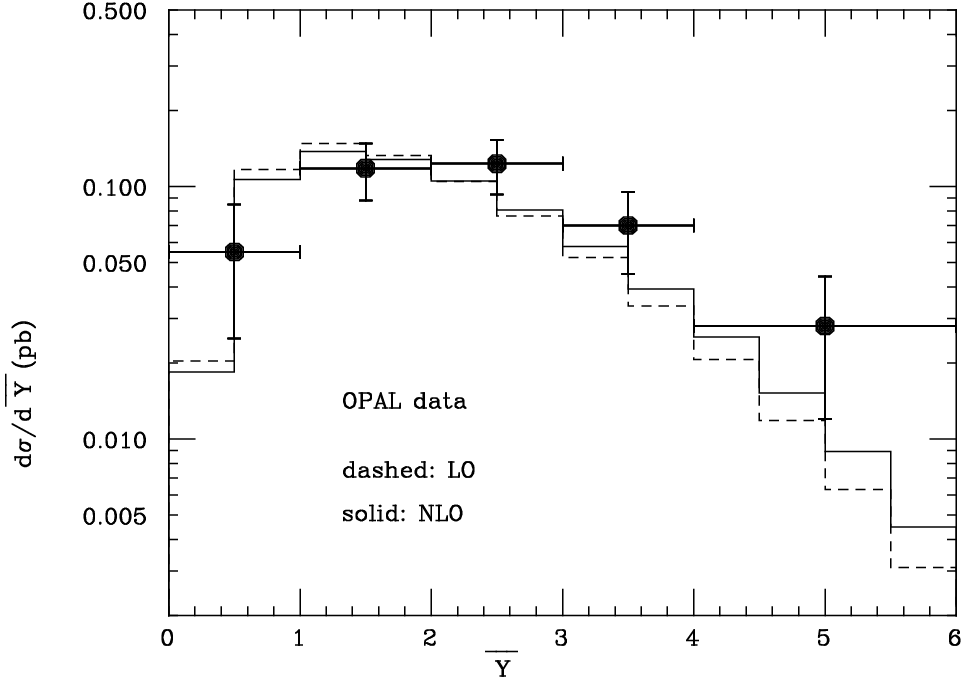,width=8cm}
\end{center}
\end{minipage}
\ccaption{}{\label{fig-opal} Differential cross sections within the OPAL
experimental cuts ($E_{1,2} > 0.4 E_{beam}$, $34 <\theta_{1,2} < 55$ mrad,
$W>5$~GeV) at $\sqrt{S} = 189 - 202$~GeV. Shown are the LO prediction
(dashed line) and the NLO one (solid). We defined~\cite{opal2000}
$Q^2=\max(Q_1^2,Q_2^2)$, and $x=Q^2/(Q_1^2+Q_2^2+W^2)$. 
$\Delta\phi$ is the difference in azimuthal angle of the outgoing 
electron and positron, measured in the $\epem$ center-of-mass frame.}
\end{center}
\end{figure}
In Fig.~\ref{fig-opal} we compare our predictions to several distributions 
related to OPAL data. A good agreement can be observed in all the
distributions, with the possible exception of the last two points in
the large-$W$ region. Where the difference between the NLO and the LO
result is somewhat more sizeable, like in the $x$ distribution and
in the large-$W$ and large-$\overline{Y}$ regions, the corrections can
be seen to change our predictions in such a way that they get closer 
to data.
\newpage

\section{Conclusions}
\label{sec:conclude}

We have calculated the NLO corrections, of $\ord(\as)$, to the
process 
\beq
\begin{array}{rcl}
e^+ + e^- & \longrightarrow & e^+ + e^- + \underbrace{\gamma^* + \gamma^*} \\
 &  & \phantom{e^+ + e^- + \gamma^*\:}\bentarrow {\rm hadrons}, 
\end{array}
\eeq
and implemented them into a Monte Carlo integrator which allows the
calculation of both total cross sections and differential distributions.
We have found the uncertainty related to the choice of renormalization 
scale to be always smaller than the net effect of including the NLO 
corrections. This means that the difference between the data and the
theoretical predictions of non-BFKL origin, which is the relevant 
quantity in any attempt to pin down signals of BFKL physics, can now 
be reliably computed at $\ord(\as)$.

When typical experimental cuts used at LEP2 by the L3 and OPAL
Collaborations are applied, NLO corrections to the total cross section 
are found to be fairly small. Larger effects can
instead be observed in the differential distributions, especially in
the regions of large $Y$ or large hadronic invariant mass $W$, where the
NLO corrections are found to increase the cross section by as much
as 50\%. No mass effects for final-state charm and bottom quarks have
been included. We recall that we have found them to decrease the LO 
cross section by 10-15\% within the set of cuts we have examined, 
thus worsening the agreement between theory and data.

When comparing to experimental results, we find good agreement with the
data measured by the OPAL Collaboration, both at the level of total cross
section and differential distributions in a number of different
observables. In this case, the effect of NLO corrections is marginal,
although the full-NLO curves are seen to be closer to data with respect to
the LO predictions: however, this comparison will become more significant 
only if the errors on data will be substantially reduced. 
Less good an agreement has instead been found when comparing to L3 data,
the NLO predictions tending to fall short of the experimental result.
In particular, when comparing with the $Y$ distribution, we can see the 
data to be sensibly higher than the theoretical prediction in the 
large-$Y$ region.

The comparison between theory and data at large $Y$'s is of course crucial, 
since a failure of fixed-order perturbative computations in describing 
the data in such a region could of course be related to the onset of 
BFKL-like effects. In this sense, no clear indication can be obtained 
from our study. If we subtract from L3 and OPAL data at large $Y$'s
our $\ord(\as)$ predictions, we get large numbers in both cases
(compared to, say, the $\ord(\as)$ results). However, while in the
case of L3 these numbers are not statistically compatible with zero,
in the case of OPAL they are statistically compatible with zero.
Thus, in order to reach a firm conclusion on this matter, the collection 
of larger statistics is unavoidable. On the other hand, if we take the 
data at their face value, there is probably an evidence of an effect
beyond NLO. It is in our opinion premature to interpret this fact in 
terms of BFKL physics. The computation of the complete $\ord(\astwo)$ 
rates would be very useful in order to understand this issue.

\section*{Acknowledgments}

We thank Jos Vermaseren for providing us with his code for the LO
computation, and Stefano Catani, Lance Dixon, Gerrit Prange and 
Maneesh Wadhwa for useful discussions. The authors thank the CERN 
Theory Division for the hospitality while this work was performed. 
This work was supported in part by the EU Fourth Framework Programme 
`Training and Mobility of Researchers', Network `Quantum Chromodynamics 
and the Deep Structure of Elementary Particles', contract FMRX-CT98-0194 
(DG 12 - MIHT), as well as by the Hungarian Scientific Research Fund 
grant OTKA T-025482.

\appendix

\section{Kinematics}
\label{sec:appa}

In this Appendix, we collect few useful formulae relevant to the
kinematics of the process we study. We define 
\beq
z_i = {2 E_i \over \sqrt{S}}\,,\qquad
\zeta_i = {Q_i^2 \over S}\equiv {-q_i^2 \over S}\,,
\label{scaled}
\eeq
with $\sqrt{S}/2$ the energy
of the incoming leptons in their center-of-mass frame.
From Eq.~(\ref{dPhiepem}) we thus get
\beq
\label{dphi}
\d\Gamma(p_{\ell_1}, p_{\ell_2}) = {S^2 \over 16(2\pi)^6}\,
\d \zeta_1\,\d \zeta_2\,\d z_1\,\d z_2\,
\d\varphi\, \d\bar{\varphi}\,.
\eeq
The azimuthal angles $\varphi$ and $\bar{\varphi}$ have to be taken in
the range $(0,2\pi)$, and it is easy to show that
\beq
0\le\zeta_i\le z_i\le 1,\;\;\;\;\;\; i=1,2\,.
\label{naivebnd}
\eeq
Using the variables defined in Eq.~(\ref{scaled}), we also get
\bea
&&
w^2(\zeta_1,\zeta_2,z_1,z_2,\varphi) = 
  (1 - z_1) (1 - z_2)
+ 2 \zeta_1 \zeta_2 - z_1 \zeta_2 - z_2 \zeta_1
\nn \\* && \qquad\qquad\qquad\qquad\;
+\,2 \cos\varphi \sqrt{\zeta_1 \zeta_2 (z_1 - \zeta_1) (z_2 - \zeta_2)}\, ,
\label{ggw}
\eea
where $w^2=W^2/S$ is the scaled squared energy of the $\gamma^*\gamma^*$
system. The requirement that $w^2>0$ further constrains $\zeta_i$, $z_i$,
and $\varphi$.

In real experimental situations the leptonic phase space is severely
restricted. The scattered leptons are observed in the forward
calorimeters, so the scattering angles $\theta_i$ off the beam 
direction are confined to a small region,
\beq
\theta_{\rm min} \le \theta_i \le \theta_{\rm max}\,,\qquad i=1,2\,.
\label{thetacuts}
\eeq
We assumed implicitly both in \eqn{ggw} and (\ref{thetacuts}) that the
$z$ axis is aligned with the incoming electron and the scattering angle
of the electron is $\theta_1$, while that of the positron is $\theta_2+\pi$. 
Typically $\theta_{\rm min}$ and $\theta_{\rm max}$ are of 
${\cal O}$(10\,mrad). Furthermore, the
energies of the scattered leptons are required to be larger than a certain
$E_\ell^{\rm min}$ in the $e^+e^-$ center-of-mass frame. In general,
$E_\ell^{\rm min} = {\cal O}$(10\,GeV) at LEP2 energies. 
In terms of the integration variables, these phase space cuts read as
\beq
{2 E_\ell^{\rm min} \over \sqrt{S}} {1 - \cos \theta_{\rm min} \over 2}
\le \zeta_i \le
{1 - \cos \theta_{\rm max} \over 2}\,,
\eeq
and $z_{\rm min} \le z_i \le z_{\rm max}$, where
\beq
z_{\rm min} = {2 \zeta_i \over 1 - \cos \theta_{\rm max}}\,,\qquad
z_{\rm max} =
\min\left(1, {2 \zeta_i \over 1 - \cos \theta_{\rm min}} \right)\,.
\eeq
In the $\epem$ center-of-mass frame, we can also use the lepton
variables in order to express the photon virtualities
\beq
- q_i^2 \equiv Q_i^2 = \sqrt{S} E_i 
\left(1 - \cos\theta_i\right),
\eeq
and the variables $y_i$, proportional to the light-cone momentum fraction 
of the virtual photon 
\beq
y_i = {q_i^0 + q_i^3 \over \sqrt{S}} = 1 - {2E_i\over \sqrt{S}} 
\cos^2{\theta_i\over 2}\,.
\label{yi}
\eeq
We also define (see Eq.(\ref{YD}))
\beq
Y=\log\frac{y_1 y_2 S}{\sqrt{Q_1^2 Q_2^2}},
\label{Ydef}
\eeq
and
\beq
\overline{Y}=\log\frac{W^2}{\sqrt{Q_1^2 Q_2^2}}.
\label{Yapproxdef}
\eeq
The variable $Y$ can also be conveniently expressed in terms of
the scaled variables defined in Eq.~(\ref{scaled}):
\beq
Y = \ln\frac{1-z_1+\zeta_1}{\sqrt{\zeta_1}}
  + \ln\frac{1-z_2+\zeta_2}{\sqrt{\zeta_2}} \:. \label{overy}
\eeq
$Y$ and $\overline{Y}$ are directly related to 
the BFKL logarithm $L$ entering Eq.~(\ref{BFKLxsec}). In fact, for large 
$W$'s the $y_i$ can be effectively interpreted as the longitudinal momentum 
fractions of the photons in the incoming leptons (since the transverse 
components of the photon momenta are much smaller than their larger light-cone 
component), and thus \mbox{$W^2\simeq y_1 y_2 S$}, which implies
\beq
Y\,\stackrel{W\to\infty}{\longrightarrow}\,\overline{Y}.
\label{YvsY}
\eeq
Furthermore (see for example Ref.~\cite{Brodsky:1997sd}), a sensible 
choice for the mass scale is 
\mbox{$\mu_{\sss\rm W}^2=c_{\sss Q}\sqrt{Q_1^2 Q_2^2}$}, 
with $c_{\sss Q}$ a suitable constant. It then follows that
\beq
L=\overline{Y}-\log c_{\sss Q}.
\label{LvsY}
\eeq

Finally, we come back to the issue of the construction of the computer
code we used to produce the phenomenological results shown in this paper.
The general strategy has been outlined in Sect.~\ref{sec:phasespace}.
As discussed there, the NLO algorithm effectively deals with the
$2\to 2$ process \mbox{$\gamma^*\gamma^*\to q\bar{q}$} (at the tree 
level and at one loop), and with the $2\to 3$ process 
\mbox{$\gamma^*\gamma^*\to q\bar{q}g$} (at the tree level). We have
to stress two important differences due to the off-shellness 
of the incoming particles ($q_i^2\ne 0$) with respect to the case described
in Refs.~\cite{Frixione:1996ms,Frixione:1997np}. Firstly,
in all the formulae given in the appendices of Ref.~\cite{Frixione:1997np},
$S$ has to be substituted with $W^2$ (the reader is urged to avoid any
confusion between the $S$ of Ref.~\cite{Frixione:1997np},
where $S$ is the
center-of-mass energy squared of the partonic system, and the $S$ used
in the rest of the present paper). Secondly, the
initial-state collinear divergences are absent. Technically, we take
this fact into account in the following way: in Eq.~(A.1) and~(A.15) of 
Ref.~\cite{Frixione:1997np}, the terms $d\sigma_{a_1a_2,i}^{(in,f)}$
and $d\sigma_{a_1a_2}^{(1,N-1r)}$ are set to zero.
Accordingly, there is no need to introduce ${\cal P}_i^{(0)}$ in the
decomposition of the ${\cal P}$ functions (see Eq.~(3.10) of
Ref.~\cite{Frixione:1997np}), and only ${\cal P}_{ij}^{(1)}$ is
non vanishing. Notice that, since now the regions of the phase space
where one of the final-state partons is collinear to one of the
initial-state particles are not infrared singular, the functions
${\cal P}_{ij}^{(1)}$ do not need to vanish in these regions.
In order to construct the code relevant to the present
paper, we implemented what discussed above in the hadronic code
of Ref.~\cite{Frixione:1997np}. On top of that, the generation
of the momenta of the leptons has been added, as discussed in
Sect.~\ref{sec:phasespace}. The matrix elements were coded
as described in Sect.~\ref{sec:lorate} and~\ref{sec:nlorate}.

\section{Notation for helicity amplitudes}
\label{sec:appb}

In order to evaluate the production rates in \sec{sec:lorate}, 
we use helicity amplitudes,
defined in terms of massless Dirac spinors $\psi_{\pm}(p)$ of fixed helicity,
\beq
\psi_{\pm}(p) = {1\pm \gamma_5\over 2} \psi(p) \equiv 
|p^\pm\rangle\, , \qquad \overline{\psi_{\pm}(p)} \equiv \langle p^\pm|
\, ,\label{spi}
\eeq
spinor products,
\beq
\langle p k\rangle \equiv \langle p^- | k^+ \rangle\, , \qquad 
\left[pk\right] \equiv \langle p^+ | k^- \rangle\, ,
\eeq
currents,
\bea
\langle i| k | j\rangle &\equiv&
\langle i^-| \slash  \!\!\! k  |j^-\rangle = 
\langle i k \rangle \left[k j\right]\, ,\nn\\
\langle i| (k+l) | j\rangle &\equiv& 
\langle i^-| (\slash  \!\!\! k + \slash  \!\!\! l ) |j^-\rangle
\eea
and Mandelstam invariants
\beq
s_{pk} = 2p\cdot k = 
\langle p k \rangle \left[kp\right]\, , \qquad t_{pkq} =
(p + k + q)^2\, .
\eeq

\end{document}